\documentclass[aps,pra,twocolumn,showpacs,superscriptaddress]{revtex4-1} 

\usepackage{tikz}
\usepackage{upgreek}
\usepackage{comment}
\usepackage{graphicx}
\usepackage{dcolumn}
\usepackage{bm}
\usepackage[T1]{fontenc}
\usepackage{epstopdf}
\usepackage{amsmath,amssymb}
\usepackage[normalem]{ulem}

\newcolumntype{L}[1]{>{\raggedright\let\newline\\\arraybackslash\hspace{0pt}}m{#1}}
\newcolumntype{C}[1]{>{\centering\let\newline\\\arraybackslash\hspace{0pt}}m{#1}}
\newcolumntype{R}[1]{>{\raggedleft\let\newline\\\arraybackslash\hspace{0pt}}m{#1}}

\newcommand{\ket}[1]    { | #1 \rangle }
\newcommand{\bra}[1]    { \langle #1 | }
\newcommand{\braket}[2] { \langle #1 | #2 \rangle }
\newcommand{\proj}[1] { | #1 \rangle \langle #1 |}
\newcommand{\matrel}[3] { \langle #1 | #2 | #3 \rangle }
\newcommand{\avg}[1]    { \langle #1 \rangle }
\newcommand{\micron}	{$\upmu$m}
\newcommand{\sub}[1]{_{\textrm{#1}}}
\newcommand{\super}[1]{^{\textrm{#1}}}
\newcommand{\m}[1]{\mathsf{#1}}
\newcommand{\f}[1]{\hat{#1}}
\newcommand{\tr}{^{\mathsf{T}}}

\newcommand{\fref}[2]{Fig.~\ref{#1}\MakeLowercase{#2}}

\newcommand{\subf}[1]{\textbf{(\MakeLowercase{#1})}}
\newcommand{\eref}[1]{Eq.~(\ref{#1})}
\newcommand{\aref}[1]{Appendix~\ref{#1}}
\newcommand{\ftit}[1]{\textbf{#1}}

\newcommand{\bchla}     {Bchl-\textit{a}}
\newcommand{\bchl}     {Bchl}

\newcommand{\mbp}[1]{{\color{blue}{#1}}}
\newcommand{\am}[1]{{\color{black}{#1}}}
\newcommand{\change}[1]{{\color{black}{#1}}}

\usepackage{layouts}
\usepackage{mathptmx}

\begin{document}

\title{Design principles for long-range energy transfer at room temperature}

\author{Andrea Mattioni}
\email{\change{Current affiliation: Department of Chemistry, School of Natural Sciences, The University of Manchester, Oxford
Road, Manchester, M13 9PL, UK.}}
\affiliation{Institut f\"ur Theoretische Physik and Center for Integrated Quantum Science and Technology IQST, Albert-Einstein-Allee 11, Universit\"{a}t Ulm, 89069 Ulm, Germany}

\author{Felipe Caycedo-Soler}
\email{\change{Deceased}}
\affiliation{Institut f\"ur Theoretische Physik and Center for Integrated Quantum Science and Technology IQST, Albert-Einstein-Allee 11, Universit\"{a}t Ulm, 89069 Ulm, Germany}

\author{Susana F. Huelga}
\affiliation{Institut f\"ur Theoretische Physik and Center for Integrated Quantum Science and Technology IQST, Albert-Einstein-Allee 11, Universit\"{a}t Ulm, 89069 Ulm, Germany}

\author{Martin B. Plenio}
\email{martin.plenio@uni-ulm.de}
\affiliation{Institut f\"ur Theoretische Physik and Center for Integrated Quantum Science and Technology IQST, Albert-Einstein-Allee 11, Universit\"{a}t Ulm, 89069 Ulm, Germany}

\begin{abstract}
{Under physiological conditions, ballistic long-range transfer of electronic excitations in molecular
aggregates is generally expected to be suppressed by noise and dissipative processes. Hence, quantum
phenomena are not considered to be relevant for the design of efficient and controllable energy transfer
over significant length and time scales. \change{Contrary to this conventional wisdom}, here we show that the robust quantum properties of
\change{small configurations of} repeating clusters of molecules can be used to tune energy transfer mechanism that take place on
much larger scales. With the support of an exactly solvable model, we demonstrate that coherent exciton
delocalization and dark states within unit cells can be used to harness
dissipative phenomena of varying
nature (thermalization, fluorescence, non-radiative decay \am{and weak inter-site correlations})
to support classical propagation over macroscopic distances.
\change{In particular, we argue that coherent delocalization of electronic excitations over just a few pigments can drastically alter the relevant dissipation pathways which influence the energy transfer mechanism, and thus serve as a molecular control tool for large-scale properties of molecular materials.}
Building on these principles,
we use extensive numerical simulations to demonstrate that they
can explain currently not understood measurements of micron-scale exciton diffusion in nano-fabricated
arrays of bacterial photosynthetic complexes.
Based on these results we provide quantum design guidelines
at the molecular scale to optimize both energy transfer speed and range over macroscopic distances in
artificial light-harvesting architectures.}
\end{abstract}

\maketitle

\section{Introduction}

{Over the last decades research into the role of coherent excitonic delocalization
in the dynamics
in photosynthetic membranes has shown that strong coherent coupling
in subunits of tightly coupled pigments can result in short-ranged excitonic delocalization in the steady state
\cite{Renger_BJ2006,Renger_2007PNAS,Novoderezhkin_BJ2006,Strumpfer2009,Strumpfer_JCP2012,Pajusalu2015}.}
Delocalization within these domains, typically restricted to individual proteins termed antenna complexes,
is essential for modeling transient and steady state optical spectra of the full {light-}harvesting ensemble.
Additionally, \change{excitonic delocalization within antenna complexes} is a crucial ingredient for a modular description
of dynamics over longer {distances} and timescales \cite{Hu_2002,Jang2004,Caycedo_2010PRL}, which, as observed
experimentally  \cite{vanAmerongen_BJ2003,Sundstrom_BJ2000,Ma_BBA2016,Flemming_JPCA2002}, 
\change{can be interpreted as a series of incoherent transfer steps, described by simple rate processes.}
These \change{energy-transfer} rates depend on the properties of the states involved, and thereby
rely heavily on  the steady state excitonic delocalization within antenna complexes, as has been shown by numerically
exact calculations \cite{Strumpfer2009,Strumpfer_JCP2012,Bennet_JPClett2018,Bennet_JPClett2018}.

\change{The modular architecture found in biological light-harvesing membranes, whereby antenna complexes containing a few pigments self-aggregate into larger structures,}
{
offers the potential for artificial solar energy conversion and molecular electronics based on such a modular
design \cite{Scholes2011, Bredas2004, Baldo2009, Aida2012}.
\change{This is made possible by} the high degree of experimental control that is
today available} for the integration of synthetic and biological structures and for the directed assembly of photosynthetic
antenna complexes isolated from living organisms \cite{Escalante2008, Escalante2010, Eisenberg2014, Vasilev2014, Harris2013, Boulais2017}
or of supramolecular dye arrays \cite{Eisele2012, Eisele2014, Akselrod2014, Clark2014, Haedler2015, Caram2016, Wan2017}.
\change{To facilitate the realization of these devices}, however, {a theoretical} understanding of the mechanisms
involved in to-date unexplained observations of \change{the large} diffusion lengths in \change{several of} these \change{light-harvesting architectures} is needed.
For {instance, the
observed micron-scale diffusion of excitations in  nano-fabricated arrays of purple bacteria antenna complexes and phycobilisome
proteins \cite{Escalante2008, Escalante2010,Eisenberg2014} exceeds} by more than one order of magnitude the
theoretical expectation for diffusion based on experimentally determined parameters under physiological conditions.
On the one hand,  this shows that an important enhancement {of} the diffusion can be achieved with hybrid technologies.
On the other hand, it stresses the necessity for theory to {establish and verify physical design principles} by which
\change{delocalization of electronic excitations over a few pigments} can enable the observed long-range energy transfer.

\change{One viable strategy for improving the diffusion lengths of excitons is to extend their lifetime, thus allowing them to propagate across longer distances.}
Recent studies on the role of dark states in solar energy conversion have provided valuable insight into the potential advantages of
protecting excitations from {losses due to} fluorescence in order to promote charge separation \cite{Creatore2013,Fruchtman2016,Wertnik2018,Hu2018,Rouse2019}.
These models, though, do not consider the microscopic origin of the interactions, thereby omitting the conditions that enable
or inhibit the active participation of dark states in the dynamics. Because dark states cannot be excited directly by light
and typically do not couple efficiently to the propagating bright states, their participation in the exciton propagation \change{is often overlooked and}
requires \change{a careful reconsideration of the typical models used to describe resonant energy transfer} \cite{Sissa2011,Liu2015}.
A further challenge is presented by the need of accounting for non-radiative decay \change{due to the interaction between electronic excitations and vibrational motion of pigment molecules}, which is responsible for the usually low
fluorescence yield of light-harvesting complexes \cite{Monshouwer1997,Helenius1997}. In fact, the competition between radiative
and non-radiative decay channels can drastically modify the dissipation landscape, leading to scenarios in which optically dark
states have a much shorter lifetime than bright states.
\am{Moreover, \change{the dynamics of excitons} can be influenced by the presence of correlations between the local vibrational environments of each pigment.
Most research on the topic has been focused on clarifying the extent to which these correlations influence the ultrafast spectroscopy of excitons, often leading to contrasting predictions
\cite{Cao2020}.
However, more relevant for our case, weak inter-site correlations can be expected to play a role on much longer timescales too and thus influence non-radiative decay.}

In the present work, we provide a theoretical model that identifies \change{the desirable features} of spectral structures and exciton
delocalization within unit cells (subunits of tightly coupled pigments) \change{that support efficient energy transfer}.
This model builds on the formation and participation of
dark states {in the dynamics} in order to achieve long-range diffusion across arrays {of} these subunits. We show how this
is possible by a combination of  excitonic delocalization within unit cells and close proximity between unit cells. Identifying each
LH2 complex as a unit cell of the transfer chain, we show that a model {that can explain} the long-range energy propagation
reported in LH2 arrays does not need to resort to previously hypothesized  \cite{Lloyd2010, Abasto2012, Yang2012} long-range quantum
coherence involving several LH2 complexes \cite{Pullerits1996, Chachisvilis1997, Monshouwer1997,Trinkunas2001,Cheng2006}.
\change{Indeed,}
we demonstrate that a theoretical description {that is} consistent with available experimental data concerning structure and optical
response of the LH2 antenna \cite{Pullerits1996, Chachisvilis1997, Monshouwer1997, Trinkunas2001, Cherezov2006} can be developed, which
reproduces the experimentally observed exciton diffusion length \cite{Escalante2010}.
\change{
This theoretical description therefore presents desirable features for transport across photosynthetic membranes.
The low-energy part of the excitonic spectrum of a unit cell comprises states that are protected against dissipation, while high-energy excitons offer fast pathways for energy transfer to neighboring units by virtue of their delocalization.
A combination of these two features allows for robust long-range energy transfer.

The remainder of this work is organized as follows. In Section II.A we present the exactly solvable model which contains the necessary features to discuss long-range exciton propagation across a modular array of excitonic unit cells. Here we discuss several energy-transfer regimes and their relation to the underlying  excitonic properties.
In Section II.B we introduce the bacterial antenna complex LH2 and provide a thorough characterization of their excitons relevant to energy transfer.
In Section II.C we present the results of our simulations of a linear array of LH2 complexes and we discuss how the mechanisms presented in Section II.A apply to a real-world application, backed by existing experimental findings.  
Section III summarizes the main results of this work and sets a broader context in which these can provide useful guidelins for energy transfer design in molecular materials.
}

\change{
\section{Model, Results and Major Insights}
}

\subsection{Minimal model of a light-harvesting array}

\begin{figure}[htb]
	\begin{center}
		\includegraphics[width=\linewidth]{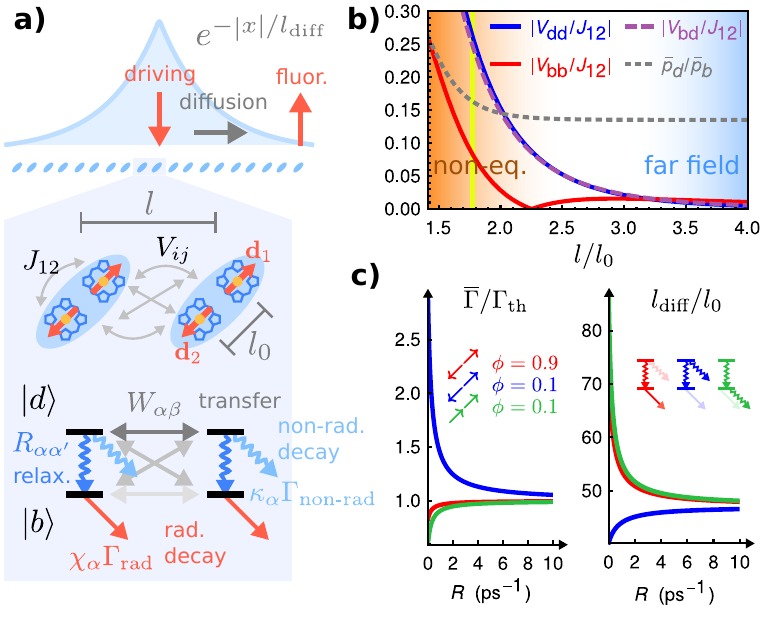}
	\end{center}
	\caption{
\ftit{Exciton propagation across a chain of dimerized units.}
\subf{a} Linear chain of dimerized unit cells separated by a distance $l$ consisting of interacting transition dipole moments, 
and sketch of the geometry and level diagram of two neighboring unit cells. Energy transfer between subunits is considered to 
be incoherent. \subf{b} Distance dependence of the couplings between unit cell excitons. For large distances (far field, $l/l_0
\gtrsim 3.2$), the coupling between bright states $V_{bb}$ dominates (red line). At short distances dark states are mainly involved 
in the energy transfer, which proceeds through the couplings $V_{dd}$ and $V_{bd}$ (blue and purple lines). Due to these couplings, 
the stationary populations $\bar{p}_b$ and $\bar{p}_d$ deviate from those in the thermal equilibrium regime that characterizes the far-field 
limit (gray dotted line). \subf{c} Effective decay rate $\overline{\Gamma}$ (left) and diffusion length $l_\text{diff}$ (right) 
as a function of the local exciton-phonon coupling, quantified by $R=\sqrt{R_{bd}R_{db}}$. The inter-dimer distance is fixed to 
$l/l_0=1.78$, corresponding to the yellow line in (a). For low fluorescence yield and anti-parallel dipoles (blue line) the coupling 
to local phonons is beneficial for long-range energy transfer as a consequence of a reduced decay rate. A larger fluorescence 
yield (red line) or a change in exciton symmetry (green line) leads to an unfavourable scaling with respect to $R$: coupling to 
local phonons hinders long-range energy transfer. The model parameters take the values $\gamma=(30 \text{ fs})^{-1}=177\text{ cm}^{-1}/\hbar$, 
$l_0=1\text{ nm}$, $d=5\text{ D}$, $\Gamma_\text{th}=(1\text{ ns})^{-1}$, $e^{-l_0/r_\text{c}}=0.9$, $k_\textrm{B}T=J_{12}=252\text{cm}^{-1}$.
}
\label{fig1}
\end{figure}

The beneficial role provided by exciton delocalization and the formation of states protected against dissipation in excitonic
energy transfer can be readily understood {in a} model system made up of dimeric unit cells (\fref{fig1}{a}). Let us
consider a homogeneous linear array of unit cells, each of them described by a Hamiltonian
\begin{equation}\label{eq1}
    \hat{H}^{(n)} = \sum_i\epsilon_i\ket{i_{n}}\bra{i_{n}}+\sum_{i\ne j}J_{ij}\ket{i_{n}}\bra{j_{n}},
\end{equation}
where the state $\ket{i_{n}}$ describes an electronic excitation of the $i$-th pigment belonging to the $n$-th unit cell.
Energy transfer across the full array is made possible by the interaction  between  pigments of different unit cells,
described by $\hat{V}^{(n,n+1)}=\sum_{i,j}V_{ij}\ket{i_{n}}\bra{j_{n+1}}+\mathrm{H.c.}$, where $V_{ij}$ are inter-dimer
dipole-dipole couplings.

If the single-pigment dephasing rate $\gamma$, the intra- and inter-dimer couplings $J_{12}$ and the inter-dimer
dipole-dipole couplings$V_{ij}$ follow
\begin{equation} \label{e.hierarchy}
    V_{ij} < \hbar\gamma < J_{12},
\end{equation}
then delocalized intra-dimer excitons defined by $\hat{H}^{(n)}\ket{\alpha_n}=E_\alpha \ket{\alpha_n}$ form. These may be used
to describe the (incoherent) energy transfer between adjacent dimers {\cite{Jang2004}.} {We refer to ``incoherent'' or ``classical''
transfer interchangeably, meaning that inter-unit cell coherences of the density operator of the full chain are negligible
($\bra{i_n}\rho\ket{j_m}\approx 0$ for $n\neq m$), which then results in classical diffusion across subunits exhibiting local
coherent dynamics.} The hierarchy of interactions in \eref{e.hierarchy} is commonly fulfilled in photosynthetic membranes and
nano-engineered arrays, where pigments  aggregate in antenna complexes within sub-nanometer distances, while the inter-complex
distances can span several nanometers, as we will discuss later in detail. For the configuration of anti-parallel transition
dipoles (i.e. $\mathbf{d}_1=-\mathbf{d}_2=d \mathbf{e}_1$)  {(indicated by red arrows in \fref{fig1}{a}),} the $n$-th dimeric
unit supports a dark (bright) exciton $\ket{d_n}$ ($\ket{b_n}$) given by the symmetric (antisymmetric) coherent superposition
$\ket{\alpha_n}$ of single pigment states, 
namely $\ket{\alpha_n}=\ket{\smash{{}^{d_n}_{b_n}}}=(\ket{1_n} \pm \ket{2_n})/\sqrt{2}$ when $\varepsilon_1=\varepsilon_2$.

Within this framework, a quantum master equation description of the full chain Hamiltonian in the presence of dephasing
and relaxation mechanisms \cite{MayKuehn} can be replaced by the classical rate equations
%
\begin{align}\label{Pauli}
\partial_t p_{\alpha}^{n}=&-\Gamma_\alpha p^{n}_{\alpha}-\sum_{\beta,m=n\pm 1} W_{\beta\alpha} p^{n}_{\alpha }+\sum_{\beta,m=n\pm 1} W_{\alpha\beta} p^{m}_{\beta}\nonumber\\
&-\sum_{\alpha'(\neq \alpha)} R_{\alpha'\alpha} p^{n}_{\alpha}+\sum_{\alpha'(\neq \alpha)} R_{\alpha\alpha'} p_{\alpha'}^{n}+I_{\alpha}^{n},
\end{align}
for the population $p^{n}_{\alpha}$ of the $\alpha$-th exciton on the $n$-th dimeric unit cell.
Here pairs of subindices $\alpha\beta$ ($\alpha\alpha'$) label excitons on different {(the same)}
unit cells. The rates that describe the transfer of excitations between unit cells $ W_{\alpha\beta}$,
their {overall} decay rate $\Gamma_\alpha$, and their thermalization rate $R_{\alpha\alpha'}$, depend on the
characteristics of the quantum states {and their environments} within these cells. Injection of excitations into the array
{can be spatially dependent and is taken to} occur with rates $I^{n}_{\alpha}$ {on site $n$}.
The rates $W_{\alpha\beta}$ can be obtained from the overlap
between homogeneous lineshape functions and depend on the
coupling matrix elements $V_{\alpha\beta}=\bra{\alpha_n}\hat{V}^{(n,n+1)}\ket{\beta_{n+1}}$  between unit cell eigenvectors
and the relative dephasing rate  $\gamma_{\alpha\beta}$ between these states  via
\begin{equation}\label{e.kpm}
\begin{aligned}
W_{\alpha\beta}=\begin{cases}
 \frac{2|V_{\alpha\beta}|^2}{\hbar^2}\frac{1}{\gamma_{\alpha\beta}} &\textrm{ \quad for $\alpha=\beta,$ } \\
\frac{2|V_{\alpha\beta}|^2}{\hbar^2}\frac{\gamma_{\alpha\beta}}{\gamma_{\alpha\beta}^2+4J_{12}^2/\hbar^2} & \textrm{ \quad otherwise},
\end{cases}
\end{aligned}
\end{equation}
as explained {in more detail} in \aref{app.lineshapes}. The dephasing rate between non-overlapping excitons
$\gamma_{\alpha\beta}=\gamma_{\alpha}+\gamma_{\beta}$ is the {sum} of the linewidths $\gamma_\alpha =
\sum_{\alpha'}R_{\alpha'\alpha}/2$, which are typically  dominated by  pure dephasing $R_{\alpha\alpha}$
over the intra-unit cell thermalization rates $R_{\alpha'\ne\alpha}$.
The thermalization rate $R_{\alpha\neq\alpha'}^{}\propto  \mathcal{J}(|E_{\alpha}-E_{\alpha'}|)\ |n(E_{\alpha}-E_{\alpha'})|$
is proportional to the phonon spectral density  $\mathcal{J}(\hbar\omega)$ and to the thermal boson occupation number
$n(\hbar\omega)$ across the excitonic manifold. {This leads to a Boltzmann distribution of the excitons when
injection, loss and} inter unit-cell transfer are {much} slower than thermalization. Excitonic pure dephasing $R_{\alpha\alpha}/2=
\sum_i |\braket{i_n}{\alpha_n}|^4\gamma \equiv \mathcal{P}^{-1}_{\alpha} \gamma $ is slower than the pigment's pure dephasing
$\gamma$ by a factor given by the inverse participation ratio $\mathcal{P}^{-1}_\alpha$  \cite{Caycedo2018}. Consequently,
excitonic delocalization ($\mathcal{P}_{\alpha}>1$) results in slower dephasing rates for unit cell excitons in comparison
to individual pigments, which, based on \eref{e.kpm}, implies an enhancement of the transfer rates between excitons in neighboring
units {with increasing} unit cell delocalization. For this dimeric system where $\mathcal{P}_{\alpha}=2$, delocalization over two
pigments of states $\ket{b_n}$ and $\ket{d_n}$  results in a two-fold speed up on the inter-dimer transfer $W_{dd}$ and $W_{bb}$,
as compared to unit cells where strong dephasing $\gamma$ or mild coupling $J_{12}$ prevents the formation of delocalized excitons
$\ket{\alpha_n}$.
Excitonic delocalization within unit cells also redistributes the optical transition dipole strength of individual pigments,
resulting in a fluorescence rate of excitons $\chi_{\alpha} \Gamma_\text{rad}$, where $\Gamma_\text{rad}$ is the single pigment
fluorescence rate and $\chi_{\alpha}=\sum_{ij}\braket{\alpha_n}{i_n}\chi_{ij}\braket{j_n}{\alpha_n}$ characterizes the optical
brightness of an exciton. The brightness quantifies the number of sites  participating in the fluorescence from exciton $\alpha$
\cite{Meier1997}, and is determined by $\chi_{ij} = (\mathbf{e}_i \cdot \mathbf{e}_j)\  j_{0}(2\pi r_{ij}/\lambda) -\frac{1}{2}\big(
\mathbf{e}_i \cdot \mathbf{e}_j - 3 (\mathbf{e}_i \cdot \mathbf{n}_{ij})  (\mathbf{e}_j \cdot \mathbf{n}_{ij}) \big)\ j_{2}(2\pi r_{ij}/\lambda)$,
where $j_\nu$ are spherical Bessel functions of the first kind, $\lambda$ is the wavelength associated to the pigment's $Q_y$
transition and $\mathbf{r}_{ij}={r}_{ij}\mathbf{n}_{ij}$ is the relative position of pigments $i$ and $j$.
In the limit $r_{ij}\ll \lambda$, the brightness reduces to the usual measure of superradiance $|\mathbf{D}_{\alpha}|^2/d^2$, i.e.
the relative dipole strength, where $\mathbf{D}_{\alpha}=\sum_i \braket{i}{\alpha} \mathbf{d}_{i}$, leading to a superradiant
$\ket{b}$ ($\chi_b\approx 2$) and a subradiant $\ket{d}$ ($\chi_d\approx0$) exciton for the dimeric unit cells of \fref{fig1}{a}.

Two ideas that will play a major role later are worth stressing at this point. First, we note that, while dark states do not couple
significantly to electromagnetic fields, they still play a central role in the energy transfer dynamics across different unit cells.
Intuitively, one expects that, if the distance between unit cells is much larger than their internal size ($l\gg l_0$), two neighboring
dimers couple via their global dipoles $\mathbf{D}_{\alpha}$. However, if the dimers are placed {sufficiently close such that their
distance becomes comparable to their spatial extent ($l\gtrsim l_0$) the dipole approximation for their mutual interaction breaks down
and states with vanishing dipole strength can start to interact via their higher moments, thus gaining some coupling strength $V_{dd}$.
This interaction can even} exceed the one of bright states $V_{bb}$ in certain configurations {as} shown in \fref{fig1}{b}, where
the coupling strengths $V_{\alpha\beta}$ are plotted against the inter-dimer distance, for the arrangement shown in \fref{fig1}{a}.
For distances $l \lesssim 3.2 l_0$, dark states couple more strongly than bright states. As a result, energy transfer through the dark
manifold can be achieved much faster than through its bright counterpart.
Secondly, we note that typical light-harvesting complexes show a rather low quantum yield of fluorescence $\phi$, i.e.  most of
the optically generated excitations are lost through non-radiative decay channels. This means that dark states, although protected
against fluorescence, could easily be more dissipative than bright states, as we show below. As discussed in \aref{app.IC} and
\ref{app.correl}, non-radiative decay rates are influenced by the presence of correlations in the vibrational environments of single
pigments. When these correlations are taken into account, the non-radiative decay rate is distributed to different excitons analogously
to what happens to the radiative rates (i.e. superradiance).
\am{In fact, static correlations between local vibrational environments can be interpreted as arising from the presence of delocalized vibrations coupling to different sites (see \aref{app.correl} for a detailed explanation). Under these circumstances, the excitonic decay rates gain a prefactor $\kappa_\alpha$ which depends on the phase with which each site contributes to the excitonic wavefunction. Depending on the specific exciton, localized excitations can interfere constructively (destructively) to yield non-radiative rates which can be larger (smaller) than their single-pigment counterpart.}
In our case, the non-radiative decay rate of the symmetric (antisymmetric) state $\ket{d_n}$ ($\ket{b_n}$) is enhanced (reduced) by
a factor $\kappa_{d (b)}=1 \pm e^{-l_0/r_\text{c}}$ with respect to its single-pigment value $\Gamma_\text{non-rad}$, where $r_\text{c}$
is the correlation length of the pigment's vibrational environment. Thus, the decay rate from exciton $\ket{\alpha_n}$ is given by
\begin{equation} \label{e.Gamma_alpha}
    \Gamma_\alpha = \chi_\alpha\Gamma_\text{rad} + \kappa_\alpha \Gamma_\text{non-rad}.
\end{equation}
Note that, for large correlation lengths, the non-radiative decay from the bright state can be significantly reduced compared to the
one from the dark state, i.e. $\kappa_b \ll \kappa_d$. Since both radiative and
non-radiative decay in light-harvesting complexes are  typically much slower than exciton thermalization, the decay of photoexcitations
in a single light-harvesting unit occurs from a thermalized exciton distribution with the average rate
\begin{equation} \label{e.gamma_th}
    \Gamma_\text{th} = \sum_\alpha p_\alpha^\text{th} \Gamma_\alpha,
\end{equation}
where $p_\alpha^\text{th} \propto e^{-E_\alpha/k_\text{B}T}$. This average decay rate is an experimentally accessible quantity, as is
the quantum yield of fluorescence
\begin{equation}
\phi = \frac{\sum_\alpha \chi_\alpha p_\alpha^\text{th}\ \Gamma_\text{rad}}{\Gamma_\text{th}}.
\end{equation}
The latter only determines the overall relative contribution of radiative and non-radiative decay, whereas the factors $\chi_\alpha$
and $\kappa_\alpha$ determine how these rates are distributed across the excitonic manifold.

So far, we have discussed how  delocalization, transition dipole geometry and environmental correlations determine the dissipation
properties of the excitonic states within unit cells, radiative and non-radiative alike,  and how the finite size of the unit cell
in densely packed arrays opens up the possibility to engage dark states into the propagation dynamics. We now have all the necessary
ingredients to determine how all these properties influence the diffusion length of excitons across the light-harvesting array. As we
are interested in  diffusion over macroscopic distances, it is useful to describe the position of a dimeric unit cell in terms of a
continuous variable $x=nl$ as $l\to 0$. Thus, bright and dark state populations at discrete sites $p^n_{\alpha}(t)$ are replaced by
the respective densities $p_\alpha(x,t)$ and \eref{Pauli} takes the form of two coupled continuous diffusion equations, which allow
for an analytical solution of the steady-state density $p_\alpha(x,\infty)$, as detailed in \aref{app.toymodel}. If we assume
that the injection and decay happen on a much slower timescale than transfer and relaxation, the solutions for local driving at $x=0$
take the simple form $p_\alpha(x,\infty) = (\bar{p}_\alpha/(2l_\text{diff})) e^{-|x|/l_\text{diff}}$, with $\bar{p}_b +\bar{p}_d=1$.
The diffusion length is given by $l_\text{diff}=l\sqrt{\overline{W}/\overline{\Gamma}}$, introducing
the effective transfer and decay rates
\begin{align}
\label{e.Weff} \overline{W}      &= \bar{p}_b (W_{bb}+W_{db}) + \bar{p}_d (W_{dd}+W_{bd}), \\
\label{e.Gamma} \overline{\Gamma} &= \bar{p}_b \Gamma_b + \bar{p}_d \Gamma_d
\end{align}
as weighted averages on the populations $\bar{p}_\alpha$.
These are determined by the ratio
\begin{equation} \label{e.pops}
    \frac{\bar{p}_d}{\bar{p}_b} = \frac{R_{db} + 2 W_{db}}{R_{bd} + 2 W_{bd}},
\end{equation}
which only depends on the rates that provide mixing between bright and dark manifolds. In particular, we observe that the presence
of {symmetric} inter-dimer bright-to-dark transfer $W_{db}=W_{bd}$ causes a deviation of the steady state populations from thermal
equilibrium $R_{db}/R_{bd}=e^{-(E_d-E_b)/k_\text{B}T}$, resulting in an increase of the stationary dark state population. Thus, the
onset of this non-equilibrium state, which can be observed at short inter-dimer separations  $l$
when the rates $W_{bd}=W_{db}$ become comparable to $R_{bd}$ and $R_{db}$ (\fref{fig1}{b}, gray dotted line),
leads to a modification of effective transfer and decay rates with respect to their equilibrium values.

In order to study the effect of dark states on energy transfer, we focus on the dependence of the diffusion length $l_\text{diff}$ on
the average intra-dimer relaxation rate $R=\sqrt{R_{bd}R_{db}}$, quantifiyng the strength of local electron-phonon coupling, at a fixed
lattice spacing $l$ such that bright-to-dark transfer and relaxation take place on a similar timescale and exciton transfer through dark
states is faster than through bright states (\fref{fig1}{b}, yellow vertical line).

First, we consider the ideal situation in which the main decay channel is radiative ($\phi=0.9$): Most of the dissipation takes place
at the low-energy bright state, which propagates more slowly than the higher-energy dark exciton. In this scenario, long range energy
transfer clearly benefits from the establishment of a non-equilibrium state with more population in the dark state. In fact, when
reducing {$R$, e.g. by decreasing the local exciton-phonon coupling,} we are {moving} population from the slow, dissipative
bright state {to} the fast, less dissipative dark state. {Furthermore, we are at} the same time reducing the effective decay
rate $\overline{\Gamma}$ and increasing the effective transfer rate $\overline{W}$, resulting in larger diffusion length $l_{\text{diff}}$
(\fref{fig1}{c}, red line).

In a more realistic scenario, we expect non-radiative decay to play a much bigger role. Let us then set $\phi=0.1$, closer to values
observed for biological light-harvesting complexes. In doing so, we also adjust the single pigment decay rates $\Gamma_\text{rad}$
and $\Gamma_\text{non-rad}$ to ensure that the effectve decay rate at thermal equilibrium $\Gamma_\text{th}$ from \eref{e.gamma_th}
remains the same as in the case just considered. Since now the main decay pathway is the non-radiative decay from the fast high-energy
exciton, an increase in the relaxation rate $R$ would lead to a longer lived excitation, but also to a slower propagation. However, the
former prevails: the reduction of effective decay $\overline{\Gamma}$ is large enough to ensure a longer-ranged propagation (\fref{fig1}{c}, blue line).
In other words, by increasing the coupling to local phonons, we counter-intuitively increase the spatial extent of energy transfer, at
the price of making it slower. This energy transfer regime can be seen as a natural extension to systems with realistic non-radiative
decay channels of the dark state protection scheme proposed in context of quantum photocells \cite{Creatore2013,Fruchtman2016,Wertnik2018,Hu2018,Rouse2019}
and recently applied to energy transfer \cite{Davidson2020}. Our generalized ``dark'' state protection mechanism makes use of
states that are protected against non-radiative decay (and are therefore ``dark''), to extend the lifetime of the excitons, which can
diffuse across longer distances.

Lastly, we observe that, while non-radiative decay pathways seem to fundamentally limit the {transport} efficiency of light-harvesting
architectures, one can work around this constraint by expoiting the symmetry of the excitonic wavefunction. Let us consider an arrangement
in which now the dipoles within each dimeric unit cell are parallel. In this situation, the intra-dimer coupling $J_{12}$ becomes negative
and the low energy exciton is now the symmetric state $(\ket{1_n}+\ket{2_n})\sqrt{2}$. While most of the optical dipole still resides in the
low energy level, now this state also becomes the most sensitive to non-radiative decay. Thus, we are left with a high-energy exciton which
shows little dissipation and fast transfer. As shown in \fref{fig1}{c} (green line), this situation is similar to the one that we considered
with $\phi=0.9$. The important difference, however, is that in this case non-radiative decay is fully taken into account, but its effect is
mitigated by exploiting excitonic delocalization within a unit cell and the ensuing redistribution of decay pathways.

\begin{figure*}[htb]
	\begin{center}
		\includegraphics[width=\linewidth]{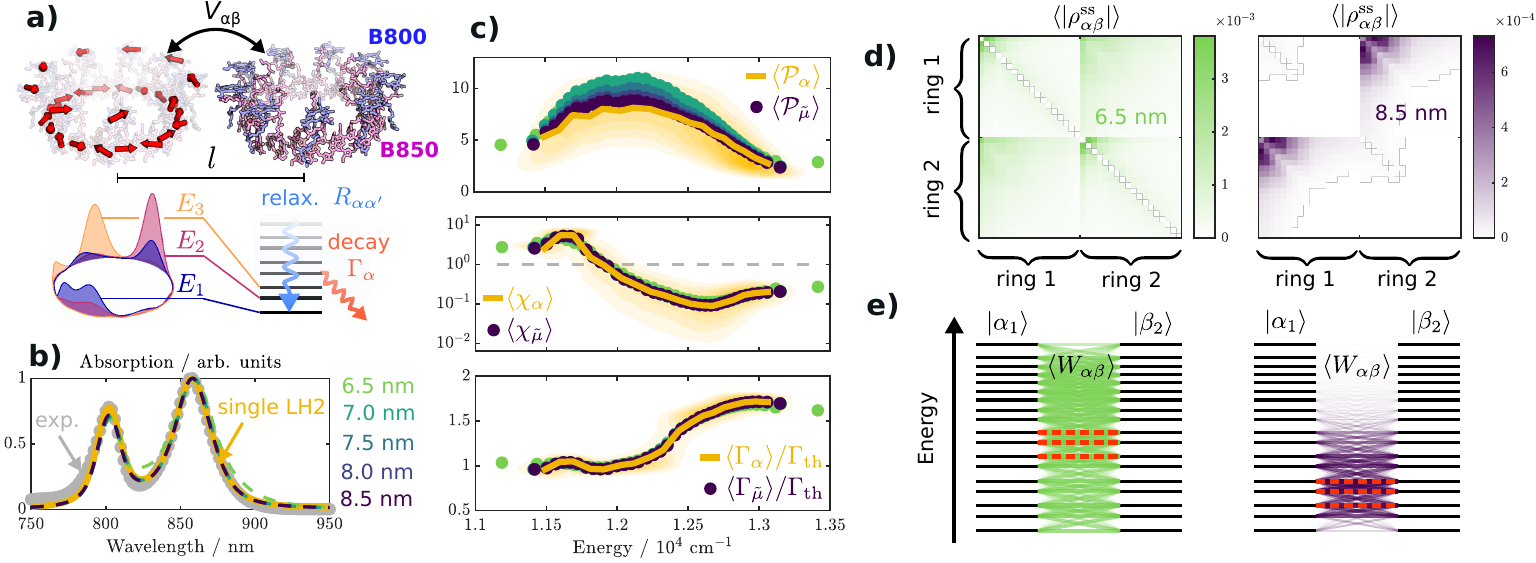}
	\end{center}
	\caption{\textbf{Excitons in bacterial light-harvesting units.}
\subf{a} Pigment arrangement in a LH2 antenna from \textit{Rps. acidophila} \cite{Cherezov2006}. B850 (violet) and B800 (blue) rings
are composed respectively {of} $18$ and $9$ BChls, whose $Q_y$ transition dipoles are indicated by red arrows. Electronic excitations
are partially delocalized and undergo relaxation within the exciton manifold, decay to the ground state or transfer
to neighboring rings.
\subf{b}
Experimental absorption {spectrum} \cite{Hildner2013} (gray dots) and theoretical fit of the B800 and B850 bands
of LH2. The inclusion of coherent inter-ring couplings in an LH2 pair (green to violet dashed lines) leads only to small deviations
from the single-ring absorption (yellow line).
\subf{c} Energy distribution of participation ratio (top), brightness (center) and decay rates (bottom) of B850 excitons, for a single ring (yellow
line) and for a coherently coupled B850 pair for different inter-ring distances (green to violet dots). For a single ring, the underlying
distribution is shown (yellow shading). Most delocalized states lie in the mid-energy range, superradiant excitons ($\chi_\alpha>1$, above
the gray dashed line) occupy the low-energy end of the spectrum, whereas the states with stronger dissipation are high-energy dark states.
Two-ring excitons are slightly more delocalized, but all two-ring quantities lie within the single-ring distributions. \subf{d} Steady state
density matrix of two coherently coupled B850 rings undergoing local relaxation and dephasing in the single-ring exciton basis, for $l=6.5\text{ nm}$
(green) and 8.5~nm (violet). Excitons are organized in ascending energy and grouped according to the ring to which they belong. Populations
are set to zero to increase contrast. \subf{e} Average energy transfer rates $\avg{W_{\alpha\beta}}$ between two B850 rings for $l=6.5\text{ nm}$
(green) and 8.5~nm (violet). Thicker and darker lines correspond to faster transfer. The three fastest transfer pathways are highlighted
(red dashed lines). For short $l$, the excitons in the mid-high energy range are mostly involved in energy transfer. Averages are performed
over $10^4$ realizations of static disorder.
}
\label{fig2}
\end{figure*}

\subsection{Bacterial light-harvesting units}

The simple dynamical model considered so far allows for analytical expressions that facilitate the identification
of the different {mechanisms on which} we base our explanation of the to-date unexplained experimental observations
of long-range energy transfer in nano-engineered arrays of LH2 photosynthetic complexes of purple bacteria \textit{Rb. sphaeroides}
\cite{Escalante2010}.
These complexes consist of a protein holding two concentric bacteriochlorophyll (\bchl{}) rings (\fref{fig2}{a}) with the
inner B850 ring consisting of 18 strongly interacting \bchl{} pigments, which at room temperature exhibit an excitonic
delocalization across about 3--6 pigments, determined by superradiance measurements \cite{Pullerits1996, Chachisvilis1997, Monshouwer1997,Trinkunas2001,Cheng2006}.
This subunit mediates the transfer between LH2 complexes under physiological conditions,
whereas the B800 pigments supports localized excitations which extend the absorption range of the LH2 complex and regulate oxidation \cite{Polivka2007}.

We consider the full B850 Hamiltonian $\hat{H}^{(n)}$
{describing} the interactions among $Q_y$ transitions of its $18$
BChls, sketched with red arrows in \fref{fig2}{a}, and study the delocalization properties of the single-ring excitons
$\ket{\alpha}$,  with $\alpha=1,\dots,18$. For realistic LH2 complexes, we need to consider different excitonic energies
$E_{\alpha}$ from realizations of pigment energies $\epsilon_i$ in order to describe the inhomogeneities (static disorder)
arising from local protein configurations. Our choice of spectral density, nearest neighbor couplings, static disorder,
geometry and magnitude of the transition dipoles are justified by previous independent analysis of experimental observations
\cite{Monshouwer1997,Stuart2011,Koolhaas1998,Hofmann2004,Jang2011,Pajusalu2015,Caycedo2018} and, when incorporated into our
model, they reproduce observed absorption spectra and superradiance enhancement as shown in \fref{fig2}{b--c} (for details
of calculation and parameters, see \aref{app.lineshapes} and \ref{app.B850}).

This parametrization of the B850 ring results
in cooperative fluorescence 
from $ \langle\sum_\alpha \chi_{\alpha}p_{\alpha}^\text{th}\rangle\approx 3$ pigments on average
(where $\avg{\cdot}$ represents the average over static disorder),
which is consistent with the experimental observations of superradiance in LH2 \cite{Monshouwer1997}.
Moreover, fixing the decay rate for an isolated LH2 complex and the quantum yield of fluorescence
to the experimentally observed values of $\Gamma_\text{th}=(1\text{ ns})^{-1}$ and $\phi=0.1$, allows
us to obtain the distribution of decay rates $\Gamma_\alpha$ across the exciton manifold. A modest
value of $r_\text{c}=5\text{ \AA}$ for the correlation radius of the pigment's vibrational environments
\cite{Renger2002,Renger_BJ2006} is sufficient to generate a significant redistribution of the dissipation
rates to the higher-energy part of the excitonic manifold (\fref{fig2}{c}), as we would expect for an
anti-parallel arrangement of neighboring dipoles.
The participation ratio shown in \fref{fig2}{c} exhibits a maximum $\langle \mathcal{P}_\alpha\rangle \approx 8$,
also consistent with previous estimates \cite{Monshouwer1997}, which underlie excitonic delocalization
constrained to small portions of the B850 ring (\fref{fig2}{a}).

We should note that for realistic LH2 center-to-center distances $l$, the aggregation into arrays does
not disrupt the excitonic manifold of single rings, as can be expected from the similarity of optical
spectra of diluted and densely packed  arrays \cite{Escalante2010}. Typical physiological conditions
and lipid-reconstituted membranes exhibit a center-to-center distance $l\approx 8$ nm, with increasing
inter-complex distances for larger lipid concentration \cite{Pflock2008,Pflock2011a,Pflock2011b}.
On the other hand, the process of nano-fabrication of LH2 arrays exploits host-guest interactions on a
nano-imprinted substrate and does not involve lipids \cite{Escalante2010}, which {allows us} to assume the
$6.2$ nm diameter of LH2 $\beta$-helices of \textit{Rb. sphaeroides} \cite{Walz1998} as the absolute minimum
for $l$.
Hence, it is reasonable to consider {center-to-center} separations of $l\gtrsim 6.5$ nm in the
nano-engineered arrays. 

To assess the robustness of the single ring excitonic manifold against the coherent interaction between neighboring LH2s,
we proceed to
diagonalize the full {two-ring} Hamiltonian $(\hat{H}^{(1)}+\hat{H}^{(2)}+\hat{V}^{(1,2)})\ket{\tilde{\mu}}= E_{\tilde\mu}\ket{\tilde \mu}$ and present in \fref{fig2}{c} the average over static disorder of their participation ratios, relative dipole strengths and dissipation rates.
Notice that even though the participation ratio in the LH2 pair $\mathcal{P}_{\tilde \mu}$ is slightly larger,  very minor changes occur in the distributions of optical brightness $\chi_{\tilde \mu}$ and dissipation rates $\Gamma_{\tilde \mu}$ with respect to the single ring eigenstates.
As a consequence, optical absorption spectra are only slightly affected by the coherent interaction between LH2 rings (\fref{fig2}{b}).
This result confirms that the coherent electronic interaction for realistic values of $l$ does not perturb significantly the excitonic structure of isolated rings. 
The robustness  of the single ring  excitonic manifold   can be understood by noticing that the  maximum coupling between any two LH2 excitons residing on different rings (averaged over static noise and relative rotations on coplanar rings) is below 40 cm$^{-1}$ even for $l=6.5\text{ nm}$,
which is much smaller than the nearest neighbor interactions within each ring $\approx 250\textrm{--}350$ cm$^{-1}$
\cite{Koolhaas1998,Caycedo2018}.
One last argument in favor of the robustness of the single ring excitonic manifold to the coherent coupling between rings
is provided by looking at the residual inter-ring coherence after exciton thermalization.
To do so, we set up a Lindblad equation describing the coherent interaction between two rings and local thermalization and dephasing,
\begin{equation}
\partial_t \rho = -\frac{i}{\hbar}[\hat{H}^{(1)}+\hat{H}^{(2)}+\hat{V}^{(1,2)},\rho]
+(\mathcal{D}^{(1)}+\mathcal{D}^{(2)})\rho,
\end{equation}
and solve for the steady state $\rho^{\text{ss}}$. Since both radiative and non-radiative decay take place on a much slower timescale,
we neglect them here.
The constant of proportionality
between the thermalization rates $R_{\alpha\alpha'}$  and the spectral density  $\mathcal{J}(|E_{\alpha}-E_{\alpha'}|)$ estimated via
fluorescence line narrowing experiments \cite{Renger2002} is such that the $\approx$ 200 fs timescale of equilibration in LH2
\cite{Stuart2011} is reproduced.
{As we are interested in demonstrating that the stationary inter-ring coherence is typically negligible, we average the
absolute value of the matrix elements $|\rho^\text{ss}_{\alpha\beta}|$ over static disorder to avoid ensemble dephasing.}
The results are shown in \fref{fig2}{d} for two inter-ring distances. Even for the shortest distance ($l=6.5\text{ nm}$),
the average steady state coherence $\avg{|\rho^\text{ss}_{\alpha\beta}|}$ between any two single-ring excitons $\ket{\alpha_1}$
and $\ket{\beta_2}$ is much smaller than $1/2$, which is the value it would take for a maximally coherent superposition of
two states $(\ket{\alpha_1}+\ket{\beta_2})/\sqrt{2}$.

At this point, we have established that {for center-to-center distances $l \ge 6.5$ nm, the incoherent energy} transfer between
neighboring B850 rings can be treated based upon the single ring eigenstates. Now we proceed to analyze the mechanisms that underlie
{this} incoherent energy transfer between neighboring LH2s.
In \fref{fig2}{e} we show how different single-ring excitons residing on two neighboring LH2 rings are connected via the average
transfer rates $\avg{W_{\alpha\beta}}$. As we saw in the previous section, a shorter separation between unit cells leads to a
substantial participation of optically dark states in the energy process, which can surpass the bright states in terms of transfer
speed. In fact, also in this case, when reducing the distance $l$ from 8.5~nm  to 6.5~nm, the fastest transfer pathways (red dashed
lines) shift from the low energy part of the spectrum, where most of the dipole strength resides, to higher-energy excitons, which
show larger delocalization.
{
This finding underlines  that the interaction in densely packed arrays does not depend on the exciton
transition dipoles but rather on their delocalization, as quantified by the participation ratio.
}

\subsection{Nano-engineered LH2 arrays}

\begin{figure*}[htb]
	\begin{center}
		\includegraphics[width=\linewidth]{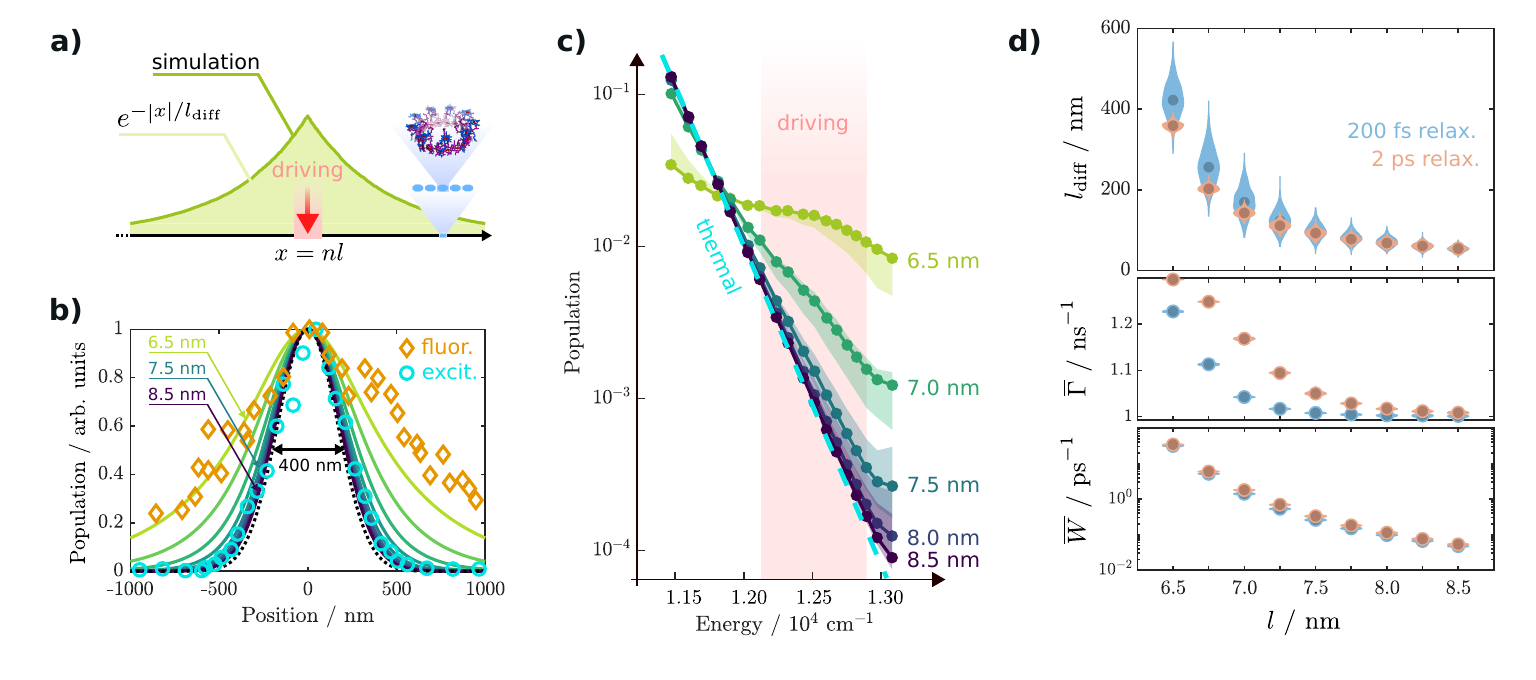}
	\end{center}
	\caption{\textbf{Exciton propagation across a chain of LH2 units.}
\subf{a} Energy transfer in a linear array of 1001 coplanar LH2 complexes subject to local driving results in an exponential population distribution. Inter-complex distances can vary between roughly \textrm{6.5 nm} in nano-engineered arrays and \textrm{8.5 nm} in biological membranes, resulting in inter-ring transfer rates $W_{\alpha\beta}\simeq (\textrm{1--10 ps})^{-1}$ in typical physiological settings.  Intra-complex dynamics is governed by inter-exciton dephasing and relaxation $R_{\alpha\alpha'}$,  with a typical time-scale of \textrm{200 fs}.
\subf{b} Spatial exciton distribution for distances $l$ ranging from 6.5 nm to 8.5 nm (green to violet lines), upon Gaussian driving with 400 nm full width at half maximum (black dotted line).
The simulations for $l=6.5\text{ nm}$ agree with the experimental data taken from \cite{Escalante2010} (circles: excitation profile; diamonds: fluorescence profile).
\subf{c} Non-equilibrium energy distribution of exciton population   averaged across the linear array for different distances $l$ (dots). 
The shaded areas represent the range in which the distribution varies when moving from the center to the ends of the array.
The injection at the central ring takes place around $12500\text{ cm}^{-1}$ from the B800 ring (pink shaded area). At small $l$, the population deviates from the Boltzmann distribution (dashed line) and transfer effectively proceeds out of thermal equilibrium.
\subf{d} Distance dependence of effective diffusion length $l_\text{diff}$ (top), decay rate $\overline{\Gamma}$ (center) and transfer rate $\overline{W}$ (bottom). Their full distributions (patches) are shown together with their average (dots).
LH2 arrays take advantage of fast local exciton thermalization (blue)
to achieve longer distance---albeit slower---energy transfer, compared to a hypothetical scenario with weaker coupling to local phonons (orange).
All averages are performed over $10^3$ realizations of static disorder.
}
	\label{fig3}
\end{figure*}

Armed with these facts, we now proceed to discuss the origin of the long-range diffusion observed in \cite{Escalante2010}.
In this experiment, simultaneous excitation  with a continuous-wave diffraction-limited laser beam and imaging of the
spatial profile of emission through confocal fluorescence detection enabled the read-out of  exciton propagation lengths
of up to 2 \micron{} in quasi-1D assemblies of LH2 complexes. To the best of our knowledge, theoretical models could only
explain such a diffusion by ignoring static disorder and underestimating dephasing \cite{Seogjoo_2001JPCB,Pajusalu2015,Renger_2018JPClett},
resulting in long-range excitonic delocalization across approximately 40 pigments \cite{Lloyd2010, Abasto2012, Yang2012},
a value that is in conflict with the experimental observations \cite{Pullerits1996, Chachisvilis1997, Monshouwer1997,Trinkunas2001,Cheng2006}.

In order to examine this experiment, we determine the rates of the Pauli master equations \eref{Pauli} for stochastic realizations
of the pigment energies $\epsilon_i$ and relative orientations of a 1D array of $1001$ coplanar LH2 complexes as shown schematically
in \fref{fig3}{A}, and study the stationary exciton distribution $\bar{p}^n_\alpha$. {For moderate excitation power which allows
us to remain in the single excitaton sector, driving can be modelled to take place on a single ring via incoherent B800-to-B850 energy
transfer (\aref{app.chain}). Exciton distributions arising from a Gaussian laser profile can then be obtained by convolution of
the driving profile with the solution for localized driving (Appendix \ref{app.toymodel}).}

Despite the presence of static disorder and the multi-level structure of each unit cell, the stationary population distribution across
the LH2 array is still characterized by an exponential distribution around the ring at which driving takes place (\fref{fig3}{a}). This
allows us to characterize the population profile by a single parameter $l_\text{diff}$. When considering an initial Gaussian beam 400 nm
wide (full width at half maximum), our model is able to reproduce the experimentally observed micron-range exciton propagation lengths
if $l=6.5\text{ nm}$ (\fref{fig3}{b}),
{while natural distances of 8.0--8.5 nm yield a barely noticeable spread of the exciton density.
This result suggests that LH2 packing density is a key factor in determining the spatial extent of energy transfer.
With a distance of $l=6.5$~nm,} 
a competition between  thermalization and transfer leads to the establishment of a non-equilibrium steady state
exciton population within antenna units $\bar{p}_{\alpha}$, that has a larger weight on high energy dark states than the thermal
distribution, as shown in \fref{fig3}{C}.
This is a clear signature that the non-equilibrium  transfer across these arrays partially proceeds via high energy dark states,
which, as {explained} above, rely on excitonic delocalization within each ring unit cell.

Despite being optically dark, these high-energy excitons are more sensitive to non-radiative decay than their low-energy bright
counterpart, as shown in \fref{fig2}{c}. Therefore, energy transfer and decay dynamics have competing effects on the diffusion
length: Local exciton-phonon interactions leading to exciton thermalization slow down the transfer of excitons by moving population
away from fast propagating high-energy states, while at the same time granting them more time to propagate, since low-energy excitons
are less sensitive to non-radiative decay.
Whether this leads to a larger diffusion length or not, depends on the specifics of the system under consideration.
{In order to test this possibility, we artificially slow down the  exciton thermalization timescale from 200 fs to 2 ps.
This corresponds in practice to forcing the steady state exciton distribution in a given ring $\bar{p}_\alpha^n$ to be
further away from thermal equilibrium $p_\alpha^\text{th}$. The shift of population towards the higher excitons results
in a shorter  ranged transfer (\fref{fig3}{d}, orange dots), revealing that these light-harvesting arrays operate in the
generalized ``dark'' state protection regime.
The relevance of such shelving mechanism is also confirmed by the behavior of the effective transfer and decay rates,
\begin{align}
\overline{W}&=\frac{1}{2}\left(
\sum_{n,\alpha,\beta}W_{\alpha\beta}^{(n,n+1)}\bar{p}_\beta^{n+1} +
\sum_{n,\beta}W_{\beta\alpha}^{(n+1,n)}\bar{p}_\alpha^{n} \right) \\
\overline{\Gamma}&=\sum_{n,\alpha}\Gamma_\alpha^{(n)}\bar{p}_\alpha^{n}.
\end{align}
(Note that the presence of static disorder in the array forces us to keep track of the unit cell index $n$ and to average
over both forward and backwards transfer rates.)
For faster relaxation, i.e. larger coupling to local phonons, both effective transfer and decay rates are significantly
decreased, as captured by \fref{fig3}{D}.
}
As noticed recently \cite{Jang2018}, the transfer rate $\overline{W}$ benefits from the engagement of dark states at small
inter-ring separations. Their participation allows for an increase in $\overline{W}$ that  exceeds the one predicted by
interactions only mediated by the collective dipoles $\mathbf{D}_{\alpha}^{(n)}$ and $\mathbf{D}_{\beta}^{(n+1)}$ of two
neighboring B850 rings. The relevance of the finite size of unit cells naturally makes the transfer more dependent on their
geometrical details and relative arrangement. Indeed, we notice for example that a systematic out-of-plane angle of just
5$^{\circ}$ as observed in lipid-reconstituted membranes \cite{Walz1998,Scheuring2003}, which slightly increases the distance
between the closest pigments in neighboring rings, slows down  the effective transfer rate and therefore decreases diffusion
length (\aref{app.chain}).
However, the apparently beneficial involvement of fast-propagating dark states in the
dynamics is countered by their sensitivity to non-radiative decay. Thus, in any situation in which we are interested in
increasing the spatial extent of energy transfer rather than its speed, we need to consider the presence of realistic
(radiative and non-radiative) decay channels and design the energy transfer process accordingly. Closely packed LH2 arrays
seem to naturally operate in a parameter regime that sustains a generalized ``dark'' state protection mechanism, where fast
exciton thermalization causes a shielding against non-radiative decay.

\am{
Finally, we notice from \fref{fig3}{d} that our model produces energy transfer rates that are in excellent agreement with existing theoretical estimates obtained for larger inter-ring separations ($l \gtrsim 7.5$~nm) \cite{Jang2018,Jang2015}, matching the conditions typically observed in reconstituted and biological light-harvesting membranes \cite{Scheuring2003,Scheuring2004}.
In this regime, inter-complex energy transfer proceeds from a completely equilibrated excitonic manifold (\fref{fig3}{c}), far from the non-equilibrium regime in which ``dark'' state shelving becomes relevant.
This suggests that the design guidelines discussed in this work, while relevant for tightly packed nano-engineered systems, might be of secondary importance for more sparsely assembled biological LH2 membranes.
}


\section{Conclusions}

In conclusion, we have shown with the help of an analytically solvable model that room temperature excitation energy
transfer can benefit from quantum dynamics within modular unit cells and demonstrated that the resulting design
principles apply in photosynthetic membranes with realistic physiological parameters, as well as in nano-fabricated
architectures.
The resulting hybrid ``quantum-classical'' design can increase both speed and propagation range thanks
to the participation of the dark states due to excitonic delocalization within unit cells. \change{On the one hand,} improved speed can be
achieved when the packing density of unit cells is made sufficiently large for the coupling of individual pigments
on different unit cells to benefit from contributions of high-lying dark states.
Crucially, close packing allows for exciton populations to depart from a thermal distribution, increasing the overall
diffusion rate via non-equilibrium energy transfer.
%
%
\change{On the other hand, the exciton propagation length is extended by intra-unit cell relaxation, biasing electronic populations towards low-energy excitons, which are less sensitive to non-radiative decay and therefore increase the overall time window over which energy transfer can take place.}
{This does not only exemplify the beneficial interplay of quantum coherent dynamics and environmental
noise \cite{plenio2008dephasing,caruso2009highly,huelga2013vibrations} but also provides basic mechanisms that
underpin} the micron-range propagation of excitations observed in artificial arrays of LH2 photosynthetic complexes.
{These can} be explained by the speed up of inter-complex transfer rates induced by dense packing of light-harvesting
units and the {protection} from non-radiative decay provided by low-energy excitons.
\am{Although we do not expect this transfer mechanism to be at play in biological LH2 membranes due to their large inter-complex separations, it could be tested on nano-engineered platforms}
{in experiments where the exciton diffusion length is measured for different light-harvesting arrays, prepared with different LH2 packing densities.}
\am{Further corroborating evidence for the mechanism that we propose is derived from experimental observation of reduced exciton lifetimes in reconstituted LH2 membranes compared to isolated complexes \cite{Pflock2008}.
Our model already captures this trend correctly
in terms of an increased population of the high-energy excitons, and thus could serve as a solid starting point for a more thorough quantitative analysis of this effect.}
\am{While the multiscale model used in this work contains all the necessary elements to discuss general energy transfer strategies in light-harvesting arrays, further improvements could be achieved by employing more refined theoretical descriptions of the light-harvesting units \cite{Jang2018rev}. We plan to do so in the future by applying some recently developed numerical techniques, allowing for an exact treatment of the non-Markovian exciton dynamics \cite{Somoza2019}.}

The fact that the speed and spatial extent of energy transfer can be directly related to excitonic delocalization
suggests the possibility of using partial delocalization restricted to single unit cells (due to the magnitude of
noise in real room-temperature  scenarios) as a resource to optimize the range of propagation of electronic excitations
in technological applications with the goal of outperforming the already extremely high efficiency of natural photosynthesis.
\am{The general nature of these design principles hints that this energy transfer scheme might find applications in a broad class of excitonic materials, not limited to the specific architecture excplicitly discussed in this work, although it will be task of future research to probe its technological feasibility.}

\section*{Acknowledgments}
The authors are grateful to A. Block and N. van Hulst for useful discussions. This work was supported by the ERC Synergy
grants BioQ and HyperQ, an IQST PhD fellowship and was made possible through the support of a grant from the John Templeton
Foundation. The opinions expressed in this publication are those of the authors and do not necessarily reflect the views
of the John Templeton Foundation.
The authors acknowledge support by the state of Baden-W\"urttemberg through bwHPC and the German Research Foundation (DFG) through grant no INST 40/575-1 FUGG (JUSTUS 2 cluster).





\renewcommand{\appendixname}{APPENDIX}
\appendix

\section{ \uppercase{Exactly solvable model}}
\label{app.toymodel}

In this appendix, we present the main steps leading to the analytical solution of the minimal model describing incoherent
energy transfer across a linear array of dimerized unit cells, which locally support quantum delocalization. We start
from \eref{Pauli} in the main text, i.e. the discrete diffusion equation of a linear array composed by dimerized unit
cells, each hosting two levels, $\ket{b_n}$ and $\ket{d_n}$ (bright and dark), which can hop to the neighboring cells
and are subject to intra-cell relaxation and fluorescence. Writing explicitly both the equations for the bright and
dark components for local injection of excitations at site $n=0$, we obtain
\begin{align}
\partial_t p_b^{n} & =  W_{bb} ( p_b^{n+1} + p_b^{n-1} - 2 p_b^{n} ) + W_{bd} ( p_d^{n+1} + p_d^{n-1} )  \nonumber \\
                  & -2 W_{db} p_b^{n} - R_{db} p_b^{n} + R_{bd} p_d^{n} - \Gamma_b p_b^{n} +I_b \delta_{n,0}\\
\partial_t p_d^{n} & =  W_{dd} ( p_d^{n+1} + p_d^{n-1} - 2 p_d^{n} ) + W_{db} ( p_b^{n+1} + p_b^{n-1} ) \nonumber \\
                  & -2 W_{bd} p_d^{n} + R_{db} p_b^{n} - R_{bd} p_d^{n} - \Gamma_d p_d^{n}  +I_d \delta_{n,0} ,
\end{align}
where $p_b^n$ ($p_d^n$) denotes the population of the bright (dark) state of the $n$-th dimer. The continuum limit is
achieved by identifying $p_\alpha^n(t)/l=p_\alpha(x,t)$ ($\alpha=b,d$) with $x=nl$ and taking the inter-dimer separation
$l$ to be vanishingly small, so that $x$ becomes a continuous variable. To simplify the discussion, we assume that the
cross-rates are equal, i.e. $W_{db}=W_{bd}=w$, which is the case for the configuration assumed in the main text. If
$W_{bd}\neq W_{db}$, a drift term in the diffusion equation is introduced, and the final exciton distribution is not
going to be symmetric around the injection point $x=0$. Thus, we obtain two coupled diffusion equations
\begin{align}
\partial_t p_b(x,t) &=  W_{bb}l^2\partial_x^2 p_b(x,t) + w l^2\partial_x^2  p_d(x,t) +I_b \delta(x)  \nonumber \\
                  &- (2w+R_{db}+\Gamma_{b}) p_b(x,t) +(2w+R_{bd})p_d(x,t)  \label{e.cont_b}  \\
\partial_t p_d(x,t) &=   wl^2\partial_x^2 p_b(x,t) + W_{dd} l^2\partial_x^2  p_d(x,t) +I_d \delta(x) \nonumber \\
                  &+ (2w+R_{db})p_b(x,t) -(2w+R_{bd}+\Gamma_{d})p_d(x,t).  \label{e.cont_d}
\end{align}

Introducing the vectors $\m{p}=(p_b,p_d)\tr$, $\m{I}=(I_b,I_d)\tr$ and the matrices
\begin{align}
\m{G}_2 &= \left( \begin{array}{cc} W_{bb} & w \\ w & W_{dd} \end{array} \right), \nonumber \\
\m{G}_0 &= \left( \begin{array}{cc} 2w+R_{db}+\Gamma_{b} & -2w-R_{bd} \\ -2w-R_{db} & 2w+R_{bd}+\Gamma_{d}  \end{array} \right),
\end{align}
we can rewrite Eq. \ref{e.cont_b}--\ref{e.cont_d} more compactly as
\begin{equation} \label{e.comp}
\partial_t \m{p}(x,t) = \big( \m{G}_2 l^2 \partial_x^2 - \m{G}_0 \big) \m{p}(x,t) + \m{I}\delta(x).
\end{equation}
Rewriting \eref{e.comp} in terms of the Fourier transform $\f{\m{p}}(q,t)=\int \textrm{d}x\ e^{-iqx}\m{p}(x,t)$ and considering the stationary state at $t\to\infty$, we obtain the solution as
\begin{equation} \label{e.solution_series}
\f{\m{p}}(q,\infty)=\frac{1}{\m{G}_0 + q^2l^2\m{G_2} } \m{I}
=\sum_{k=0}^\infty (-q^2l^2 \m{G}_0^{-1} \m{G}_2 )^k \m{G}_0^{-1}\m{I}.
\end{equation}
The solution can be brought to a much more transparent form if we assume that the decay described by $\Gamma_\alpha$ takes place on a much slower timescale than inter- and intra-dimer energy transfer.
Within this approximation, which is typically satisfied in light-harvesting complexes, we have
\begin{equation}
\m{G}_0^{-1}=\frac{1}{\overline{\Gamma}}\left( \begin{array}{cc} \bar{p}_b & \bar{p}_b \\ \bar{p}_d & \bar{p}_d \end{array} \right)
\end{equation}
to leading order in $\Gamma_\alpha$, where $\overline \Gamma$ and $\bar{p}_\alpha$ were defined in \eref{e.Gamma} and \eref{e.pops}.
This simple form allows us to sum the series in \eref{e.solution_series} to
\begin{equation}
 \label{e.solution_fourier}
\f{\m{p}}(q,\infty)=\frac{I_b+I_d}{\overline \Gamma} \frac{1}{1 + q^2 l^2 \overline{W}/\overline{\Gamma} }
\left( \begin{array}{c} \bar{p}_b \\ \bar{p}_d \end{array} \right),
\end{equation}
where $\overline W$ was defined in \eref{e.Weff}.
Taking the inverse Fourier transform leads to the final form of the solution
\begin{equation} \label{e.solution}
\m{p}(x,\infty)=\frac{I_b+I_d}{\overline \Gamma} \frac{1}{2 l_\text{diff}}
e^{-|x|/l_\text{diff}}
\left( \begin{array}{c} \bar{p}_b \\ \bar{p}_d \end{array} \right),
\end{equation}
which is the exciton distribution discussed in the main text. As one would expect, if inter-exciton conversion 
due to the cross-rates $R_{bd}$, $R_{db}$ and $w$ is much faster than the decay timescale, both bright and dark 
manifold have the same final distribution, differing only by the a normalization constant $\bar{p}_\alpha$. It 
is easy to check that we would get the same exponential distribution if we considered a diffusion process involving 
unit cells containing a single level rather than two, with transfer and dissipation rates $\overline{W}$ and 
$\overline{\Gamma}$. Therefore, a dimeric unit cell results in the same type of diffusion as a monomeric one, 
where the large-scale diffusion properties can be tuned by changing the local dimer parameters.

A completely analogous procedure leads to the solution of the case in which $W_{bd}\neq W_{db}$, which we do not present in detail here.
The imbalance between these two rates does not change the definitions of $\bar{p}_\alpha$, $\overline W$ and $\overline \Gamma$ (which 
only depend on the average cross-rate $w=(W_{bd}+W_{db})/2$), but leads to a drift coefficient proportional to $\Delta=(W_{bd}-W_{bd})(\bar{p}_b-\bar{p}_d)$.
As a final result, the exciton distribution is still exponentially localized around the injection point $x=0$, but the diffusion lengths 
for $x<0$ and $x>0$ are different. Thus, the diffusion is not symmetric. As a possible application, this property might be used to design 
excitonic wires that are able to switch the dominant direction of diffusion by small changes of their unit-cell properties, for example 
implementing artificial photoprotection.

Although in the case of the exactly solvable  model we consider a completely homogeneous system, it is possible
to show that the presence of moderate static {disorder does} not modify the exponential shape
of the stationary exciton distribution, but only leads to a reduction of the diffusion length.
This explains why, also in the case of the LH2 array presented in the main text, we still obtain
an exponential distribution. To grasp of the effects of static disorder, we consider a linear array
of monomeric unit cells with some slight random inhomogeneity in the transfer rates. The transfer
rate from site $n$ to site $n+1$ (and vice versa) is given by $W_{n} = \overline{W} + \delta W_n$,
where we assume $\avg{\delta W_n}=0$ and $\avg{\delta W_n \delta W_m}=\sigma^2 \delta_{nm}$. The
average population distribution $\avg{p(x)}$ for small disorder can be obtained by expanding the
solution in the form of \eref{e.solution_series} to second order in $\delta W_n$ and taking the
ensemble average. This leads to an exponential average exciton distribution with diffusion length
$l_\text{diff}'=l_\text{diff}(1-\sigma^2/2\overline{W}^2)$, independent of the specific distribution
of the rate fluctuations $\delta W_n$.

To conclude this section, we note that the stationary solution $\m{p}(x,\infty)$ for local exciton
injection $\delta(x)$ is sufficient to determine the exciton profile $\m{p}'(x,\infty)$ for any other
(normalized) injection  profile $g(x)$. In fact, defining $\f g (q)$ the Fourier transform of $g(x)$,
we immediately obtain that the new solution satisfies
\begin{equation}
    \f{\m{p}}'(q,\infty)=\frac{\f g (q)}{\m{G}_0 + q^2l^2\m{G_2} } \m{I}
    =\f{\m{p}}(q,\infty) \f g (q).
\end{equation}
Transforming back to real space, we  obtain the new solution for generic driving as a convolution between the solution for local driving and the generic driving profile, namely
\begin{equation}
\m{p}'(x,\infty)=
\int \text{d}x'\ \m{p}(x',\infty) g (x-x').
\end{equation}
The same principle allows us to draw conclusions for exciton propagation in a LH2 array upon driving with a Gaussian laser profile, using only the results of simulations for local driving.

\section{ \uppercase{ Microscopic origin of non-radiative decay } }
\label{app.IC}

In this appendix, we derive a microscopic Hamiltonian describing vibrationally-induced non-radiative decay of a photoexcited molecular state.
We start from the \textit{ab initio} Hamiltonian of a single chromophore.
We determine the form of the non-adiabatic coupling at the basis of nonradiative decay in the usual harmonic approximation for intra-molecular vibrations. We use this result to compute the internal conversion rate for delocalized excitonic states of a molecular aggregate.

\subsection{Molecular Hamiltonian}

The full Hamiltonian of a molecule {is given} by \cite{MayKuehn}
\begin{align} \label{e.H}
H({\bf r},{\bf p};{\bf R},{\bf P})
&=T_\text{el}({\bf p})
+V_\text{el-el}({\bf r})
+V_\text{el-nuc}({\bf r};{\bf R}) \\
&+V_\text{nuc-nuc}({\bf R})
+T_\text{nuc}({\bf P}), \nonumber
\end{align}
where $T$ denotes the kinetic energy and $V$ the Coulomb interactions. The sets of electronic coordinates 
and momenta are denoted by ${\bf r}$ and ${\bf p}$, whereas the nuclear degrees of freedom are described 
by ${\bf R}$ and ${\bf P}$. The typical challenge in condensed matter physics is to find approximate eigenstates 
and eigenvalues of this Hamiltonian. The huge difference between electronic and nuclear masses justifies
the typical Born-Oppenheimer (BO) approach, where the electronic degrees of freedom are treated on a fully 
quantum level (i.e. $\mathbf{r}\to \hat{\mathbf{r}}$ and $\mathbf{p}\to \hat{\mathbf{p}}$), whereas the 
nuclear coordinates are kept fixed and their momenta are initially neglected. This allows to find the 
adiabatic electronic eigenstates by diagonalization of the electronic Hamiltonian  at fixed $\bf R$:
\begin{multline} \label{e.adiabatic}
[ T_\text{el}(\hat{\bf p})+V_\text{el-el}(\hat{\bf r})
+V_\text{el-nuc}(\hat{\bf r};{\bf R}) \\ +V_\text{nuc-nuc}({\bf R}) ]
\ket{\phi^i_\mathbf{R}}
= \varepsilon^i_{\bf R} \ket{\phi^i_\mathbf{R}}.
\end{multline}
The eigenstates $\ket{\phi^i_\mathbf{R}}$ and eigenenergies $\varepsilon^i_{\bf R}$---which take the name of potential energy surfaces (PES)---depend parametrically on the nuclear coordinates.
Now we express the Hamiltonian \eqref{e.H} in terms of these adiabatic electronic states.
Although their interpretation as physically relevant states depends on whether the nuclear kinetic energy is negligible or not, they always form a perfectly legitimate orthonormal basis of the electronic Hilbert space at fixed $\bf R$.
Making use of \eqref{e.adiabatic}, we see that
\begin{equation} \label{e.Hij}
\bra{\phi^i_\mathbf{R}} H(\hat{\bf r},\hat{\bf p};{\bf R},{\bf P}) \ket{\phi^j_\mathbf{R}}
= \delta^{ij} \varepsilon^i_{\bf R}
+ \bra{\phi^i_\mathbf{R}} T_\text{nuc}({\bf P}) \ket{\phi^j_\mathbf{R}}.
\end{equation}
Note that the second term on the right-hand side does not necessarily reduce to $\delta^{ij} T_\text{nuc}({\bf P})$, due to the parametric dependence on $\bf R$ of the electronic states.
However, if nuclei are considered as classical particles, the nuclear kinetic energy does not affect the electronic adiabatic wavefunction in any way, and the full molecular Hamiltonian \eqref{e.H} takes the well-known BO form
\begin{equation} \label{e.Hclassical}
\hat{H}_\text{BO}({\bf R},{\bf P}) = \sum_i \proj{\phi^i_{\bf R}} \left( T_\text{nuc}({\bf P}) + \varepsilon^i_{\bf R} \right).
\end{equation}
The adiabaticity of this Hamiltonian is represented by the fact that the nuclear motion does not couple different electronic states: when the electrons are in the state $\ket{\phi^i_\mathbf{R}}$, the nuclei evolve according to the effective Hamiltonian $h_i(\mathbf{R},\mathbf{P})=T_\text{nuc}({\bf P}) + \varepsilon^i_{\bf R}$.
Introducing a diagonal nuclear mass tensor $\m M$ and expanding the $i$-th PES for small deviations 
from its stable equilibrium configuration ${\bf R}^i$ (i.e. close to its minimum $\varepsilon^i_{{\bf R}^i}$), the nuclear Hamiltonian $h_i$ becomes
\begin{align} \label{e.hi}
h_i({\bf R,P})
& =\varepsilon^i_{{\bf R}^i}+\frac 1 2 \left[ {\bf P}\tr{}{\m M}^{-1}{\bf P}
+ ({\bf R}-{\bf R}^i)\tr{} {\m H} ({\bf R}-{\bf R}^i)  \right] \nonumber \\
& =\varepsilon^i_{{\bf R}^i}+\frac 1 2 \left[ {\bm \pi}\tr{}{\bm \pi} + ({\bm \xi} - {\bm \xi}^i)\tr{} {\m D} ({\bm \xi} - {\bm \xi}^i)  \right],
\end{align}
where $\m H \geq 0$ is the Hessian matrix of the PES $ \varepsilon^i_{\bf R} $ at ${\bf R}^i$.
On the second line, we introduced the mass-rescaled normal coordinates and momenta
${\bm \xi}=\m{U}\m{M}^{1/2}{\bf R}$ and
${\bm \pi}=\m{U}\m{M}^{-1/2}{\bf P}$,
where  $\m{U}$ is the unitary transformation that diagonalizes the mass-rescaled Hessian, i.e.
$ \m{U}\tr{} \m{D} \m{U} = \m{M}^{-1/2}\m{H}\m{M}^{-1/2}$. $\m D$ has only diagonal entries, corresponding to the square of the normal frequencies $\omega_k^2$.
As usual, we have neglected the Duschinsky mixing, that is,  we have assumed that the Hessian  $\m H$ at the nuclear equilibrium configuration is independent of the specific PES, so that it is diagonalized by the same transformation and yields the same eigenvalues (i.e. vibrational frequencies) for the PESs in which we are interested.

At this point, we can easily quantize the nuclear normal modes by
redefining them in terms of a set of bosonic annihilation (creation) operators $\hat{b}_k$ ($\hat{b}_k^\dagger$). Assuming from now on $\hbar=1$, we have
\begin{align}
\hat{\xi}_k & = \frac{1}{\sqrt{2\omega_k}} (\hat{b}_k+\hat{b}_k^\dagger), \\
\hat{\pi}_k & = -i \sqrt{ \frac{\omega_k}{2} } (\hat{b}_k-\hat{b}_k^\dagger).
\end{align}
The quantized vibrational Hamiltonian \eqref{e.hi} becomes
\begin{equation} \label{e.hq}
\hat{h}_i = \varepsilon^i_{{\bf R}^i} + \sum_k \omega_k \left[ (\hat{b}_k-\sqrt{s_k^i})^\dagger(\hat{b}_k-\sqrt{s_k^i}) + \frac{1}{2} \right].
\end{equation}
Note that we have introduced the Huang-Rhys factors $s_k^i={\xi_k^i}^2 \omega_k/2$, which in general depend on the PES under consideration.

Since our ultimate goal is to describe the dynamics of photo-excitations, we focus on two electronic states $\ket{\phi^g_{\bf R}}$ and $\ket{\phi^e_{\bf R}}$, describing the electronic ground state and the first optically excited state.
We neglect any further dependence of these states on the nuclear coordinates and refer to them simply as $\ket{\phi_g}$ and $\ket{\phi_e}$.
If we refer all energies and nuclear coordinates to the minimum of the PES of the electronic ground state, the BO Hamiltonian \eqref{e.Hclassical} takes the  usual spin-boson form with pure dephasing interaction
\begin{equation} \label{e.spin-boson}
\hat{H}_\text{BO} = \varepsilon \proj{\phi_e}
+\proj{\phi_e}\sum_k g_k \left( \hat{b}_k + \hat{b}^\dagger_k \right)
+\sum_k \omega_k \hat{b}^\dagger_k \hat{b}_k,
\end{equation}
where $\varepsilon = \varepsilon^e_{{\bf R}^e} + \sum_k s_k\omega_k$ and $g_k = -\omega_k \sqrt{s^e_k}$.
The adiabatic electron-phonon coupling can then be described by the spectral density
\begin{equation} \label{e.SD}
\mathcal{J}(\omega)=\sum_k g_k^2 \delta(\omega-\omega_k).
\end{equation}

\subsection{Non-adiabatic electron-phonon coupling}

Now, we turn back to \eqref{e.Hij} and consider the effect that the nuclear kinetic energy has on the adiabatic electronic wavefunction.
For this, we take into account the quantum nature of nuclei as well, i.e. considering ${\bf R}\rightarrow \hat{\bf R}$ and ${\bf P}\rightarrow \hat{\bf P}$ as quantum mechanical operators.
Note that the second term on the right-hand side of \eqref{e.Hij} acts non-trivially on both electronic and nuclear degrees of freedom.
To better understand this, we consider how the nuclear kinetic energy operator acts on a product state  formed by  $\ket{\phi^j_\mathbf{R}}$ and a nuclear state $\ket{\chi_\nu}$.
In coordinate representation, $\hat{\bf P}$ becomes the differential operator $-i\partial_{\bf R}$, therefore we have
\begin{multline} \label{e.Tcoord}
\bra{{\bf r},{\bf R}} T_\text{nuc}(\hat{\bf P}) \ket{\phi^j_\mathbf{R}, \chi_\nu}
 = \sum_k \frac{\hat{P}^2_k}{2M_k} {\phi^j_\mathbf{R}}({\bf r}){\chi_\nu}({\bf R}) \\
 = \sum_k \frac{1}{2M_k}\left[ \phi_{\bf R}^j({\bf r})\left(\hat{P}_k^2 {\chi_\nu}({\bf R})\right)
       + 2\left(\hat{P}_k\phi_{\bf R}^j({\bf r})\right)\left(\hat{P}_k {\chi_\nu}({\bf R}) \right) \right. \\ \left. + \left(\hat{P}_k^2 \phi_{\bf R}^j({\bf r})\right){\chi_\nu}({\bf R}) \right]
\end{multline}
Using \eqref{e.Tcoord}, we can now compute the matrix element of the nuclear kinetic energy operator between two arbitrary electronic-vibrational states as
\begin{widetext}
\begin{align} \label{e.Tmatrel}
\bra{\phi^i_{\bf R},\chi_\mu} T_\text{nuc}(\hat{\mathbf{P}}) \ket{\phi^j_{\bf R},\chi_\nu}
 & = \int\text{d}{\bf r}\int\text{d}{\bf R}\ \braket{\phi_{\bf R}^i,\chi_\mu}{{\bf r},{\bf R}}
     \bra{{\bf r},{\bf R}} T_\text{nuc}(\hat{\bf P}) \ket{\phi^j_\mathbf{R},\chi_\nu} \\
 & = \delta^{ij} \bra{\chi_\mu} T_\text{nuc}(\hat{\bf P}) \ket{\chi_\nu}
  -i \sum_k \frac{1}{M_k}\int\text{d}{\bf R}\ \braket{\phi_{\bf R}^i}{\partial_{R_k}\phi_{\bf R}^j}
  \chi_\mu^*({\bf R})\hat{P}_k{\chi_\nu}({\bf R})
 - \sum_k \frac{1}{2M_k}\int\text{d}{\bf R}\ \braket{\phi_{\bf R}^i}{\partial^2_{R_k}
 \phi_{\bf R}^j} \chi_\mu^*({\bf R}){\chi_\nu}({\bf R}) \nonumber
\end{align}
\end{widetext}
where the three terms appearing on the second line of \eqref{e.Tmatrel} directly originate from those appearing in \eqref{e.Tcoord}.
It is easy to spot the first of them as the one giving rise to the BO Hamiltonian \eqref{e.Hclassical}.
On the other hand, the terms involving derivatives of the adiabatic electronic wavefunction with respect to the nucrear coordinates are not necessarily proportional to $\delta^{ij}$, and therefore describe the non-adiabatic coupling between different PESs.
Within the BO framework, they are neglected by merit of the argument that the electronic wavefunctions depend very weakly on the nuclear coordinates, therefore their derivatives are negligible.
In fact, electronic wavefunctions usually vary significantly over distances of 1--10 \AA{}, corresponding to typical inter-nuclear separations.
On the other hand, typical nuclei in organic molecules jiggle by about 0.01--0.1 \AA{} around their equilibrium position.
On this length-scale, therefore, electronic wavefunctions are expected to be fairly smooth and almost constant.
Thus, the last two terms in \eqref{e.Tmatrel} can be regarded as higher order contributions of a perturbative expansion whose zeroth order term corresponds to the BO Hamiltonian.

For simplicity, let us focus only on the lowest non-zero order of this non-adiabatic perturbation, i.e. the one involving the first derivative of the adiabatic electronic wavefunction.
Let us define the quantity
$A^{ij}_k({\bf R}):=-i \braket{\phi^i_{\bf R}}{\partial_{R_k}\phi^j_{\bf R}}$, which is closely related to the Berry connection.
Note that we can always assume the phase of the adiabatic electronic wavefunctions to be independent of ${\bf R}$, corresponding to a specific gauge choice for the Berry connection.
This, together with the normalization condition $\braket{\phi^i_{\bf R}}{\phi^i_{\bf R}}=1$, immediately implies that $F^{ii}_k({\bf R})=0$, meaning that this term only connects different electronic states and does not modify the vibrational Hamiltonian within a given PES.
On top of that, the orthogonality condition $\braket{\phi^i_{\bf R}}{\phi^j_{\bf R}}=0$ further imposes a hermiticity constraint
${A^{ij}_k({\bf R})^*} = A^{ji}_k({\bf R})$.
As usually assumed for optical transitions, we neglect the dependence of the coupling between different PESs on the nuclear configuration.
This means that $A^{ij}_k({\bf R})\approx A^{ij}_k({\bf R}^i)$ is practically constant in the integral over $\bf R$ in \eqref{e.Tmatrel}.
We also neglect any further dependence of the adiabatic states on the nuclear coordinates, as we did in the previous section in order to recast the BO Hamiltonian \eqref{e.Hclassical} into spin-boson form \eqref{e.spin-boson}.
Thus, we can finally write the initial molecular Hamiltonian \eqref{e.H} including the lowest order non-adiabatic correction as $ \hat{H}_\text{BO} + \hat{H}_\text{non-ad} $, where
\begin{align} \label{e.Hmonomer}
\bra{\phi_i}\hat{H}_\text{non-ad}\ket{\phi_j} & =
\sum_k A^{ij}_k({\bf R}^i) \frac{\hat{P}_k}{M_k} \\
 & = -i \sum_k \alpha^{ij}_k \sqrt{\frac{\omega_k}{2}} (\hat{b}_k - \hat{b}_k^\dagger) \nonumber,
\end{align}
where in the last step we have rewritten the nuclear momenta in terms of the normal modes defined above and introduced the transformation
${\bm \alpha}^{ij} = \m{U}\m{M}^{-1/2}\mathbf{A}^{ij}({\bf R}^i)$
on the coefficients $A^{ij}_k({\bf R}^i)$.

Thus, the non-adiabatic electron-phonon coupling can lead to transitions between different electronic states $\ket{\phi_i}$ and $\ket{\phi_j}$ initiated by the action of nuclear momenta.
If we consider only two electronic states, namely the ground state $g$ and one optically excited state $e$, we can see this as causing non-radiative transitions between $g$ and $e$ mediated by the coupling constant
$f_{k}:=\alpha^{ge}_k\sqrt{\omega_k/2}$.
This process goes under the name of internal conversion (IC), and is therefore determined by the following spectral density
\begin{equation} \label{e.icsd}
\mathcal{J}_\text{non-ad}(\omega)=\sum_k f_k^2 \delta(\omega-\omega_k),
\end{equation}
analogous to \eqref{e.SD}.

\subsection{Internal conversion in molecular aggregates}

While it is generally true that the non-adiabatic couplings are much weaker than the pure-dephasing electron-phonon coupling, they still play a role on longer time-scales, where other phenomena such as fluorescence enter into play.
Therefore we should devise a way to treat IC on the same footing as fluorescence in molecular aggregates.
We consider an aggregate of $N$ interacting two-level chromophores, each one described by the Hamiltonian \eqref{e.Hmonomer}.
We focus on the subspace spanned by the global ground state
$\ket{g}=\ket{\phi_g}_1\dots\ket{\phi_g}_N$
and the single excitations
$\ket{i}=\ket{\phi_g}_1\dots\ket{\phi_e}_i\dots\ket{\phi_g}_N$.
Assuming that the only site dependent parameters of our model are the electronic excitation energies $\varepsilon_i$ and couplings $J_{ij}$, the Hamiltonian of the aggregate is
\begin{align} \label{e.Haggr}
\hat{H}_\text{aggr} & = \sum_i \varepsilon_i \proj{i} + \sum_{i\neq j} J_{ij}\ket{i}\bra{j} + \sum_i\sum_k \omega_k \hat{b}_{ik}^\dagger\hat{b}_{ik} \nonumber \\
 & \quad +\sum_{i}\proj{i}\sum_k g_k \left( \hat{b}_{ik}+ \hat{b}_{ik}^\dagger \right) \nonumber \\
 & \quad  -i \sum_{i}\left(\ket{g}\bra{i}+\ket{i}\bra{g}\right)\sum_k f_k  (\hat{b}_{ik}-\hat{b}_{ik}^\dagger).
\end{align}
On the first line, we recognize the free Hamiltonian of the excitonic system (diagonalized by excitons $\ket{\alpha}=\sum_i\ket{i}\braket{i}{\alpha}$ with energy $E_\alpha$) and the free Hamiltonian of environmental vibrations.
On the second and third line, we find the system-bath interactions, mediated by the two bath operators
\begin{align}
\hat{G}_i & =    \sum_k g_k \left( \hat{b}_{ik} + \hat{b}_{ik}^\dagger \right),\label{e.G}\\
\hat{F}_i & = -i \sum_k f_k \left( \hat{b}_{ik} - \hat{b}_{ik}^\dagger \right),\label{e.F}
\end{align}
which respectively cause pure dephasing in the site basis and non-radiative transitions between site excitations and global ground state.
Postponing a more detailed justification for later, we allow for correlations between different vibrational environments to be present when the bath is at equilibrium,
\begin{align}
\avg{\hat{G}_i (t) \hat{G}_j (0) }_\text{eq} & = \kappa_{ij} \mathcal{G}(t), \label{e.GGeq} \\
\avg{\hat{F}_i (t) \hat{F}_j (0) }_\text{eq} & = \kappa_{ij} \mathcal{F}(t), \label{e.FFeq}
\end{align}
where the time evolution is computed with respect to the free bath Hamiltonian and the expectation value $\avg{\cdot}_\text{eq}$ is taken on the stationary state of the bath.
The correlation functions $\mathcal{G}(t)$ and $\mathcal{F}(t)$ are assumed to be site-independent, and only depend on the temperature of the vibrational bath  and on the spectral densities
$\mathcal{J}(\omega)$ and $\mathcal{J}_\text{non-ad}(\omega)$.
We assume that the internal timescale of the bath is sufficiently fast and the system-bath coupling is sufficiently weak so that the dynamics of electronic excitations can be described by a Lindblad equation in the exciton basis.

We focus on internal conversion for now.
Following the microscopic derivation outlined in \cite{rivas2011open},
we obtain a Lindblad equation with
jump operators
$\ket{g}\bra{\alpha}$
and rates
\begin{equation} \label{e.kappa_alpha}
\kappa_\alpha\Gamma_\text{non-rad}=\sum_{i,j}\braket{\alpha}{i} \kappa_{ij} \braket{j}{\alpha}\Gamma_\text{non-rad},
\end{equation}
appearing in \eref{e.Gamma_alpha}.
The rate $\Gamma_\text{non-rad}$ is the single-pigment non-radiative rate, determined by
\begin{align} \label{e.ICrate}
\Gamma_\text{non-rad} &= \int\text{d}t\ e^{i\omega t} \mathcal{F}(t) \\
 &=2\pi \mathcal{J}_\text{non-ad}(|\omega|)|n(\omega)+1|. \nonumber
\end{align}
In principle, it depends on the frequency at which the specific transition to the ground state takes place (i.e. $\omega=E_\alpha$ for  $\ket{g}\bra{\alpha}$).
However, since optical frequencies are much higher than vibrational and thermal energies,
we can assume $\mathcal{J}_\text{non-ad}(\omega)$ to be fairly small and constant across the excitonic energies $E_\alpha$.

\begin{figure}[htb]
	\begin{centering}
		\includegraphics[width=\linewidth]{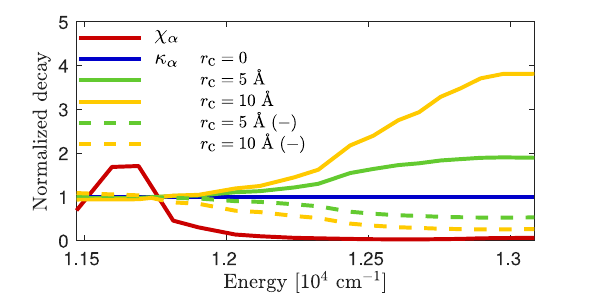}
		\caption{{\bf Excitonic decay rates. } Normalized decay rates from the excitonic manifold to the ground state for different processes
		in the B850 band of LH2:
		radiative (red), non-radiative with increasing bath correlation length (blue, green and yellow solid lines).
		The dashed lines correspond to models with the same correlation length, but assuming anticorrelated bath
		fluctuations for couples of pigments that have opposite transition dipole moments.
		The decay rates are normalized such that on average they all give the same decay rate from a thermal exciton distribution.
		The results are obtained as ensemble averages over $10^4$ realizations of static disorder.
		}
		\label{f.decay}
	\end{centering}
\end{figure}

The correlations introduced in \eqref{e.GGeq} and \eqref{e.FFeq} are crucial for the determination of the distribution of the IC rate across the excitonic manifold, as shown in Fig. \ref{f.decay}.
For example, if the correlations are absent, i.e. $\kappa_{ij}=\delta_{ij}$, Eq. \eqref{e.ICrate} predicts that the IC rate is essentially the same for all excitons (blue line).
If, instead, we allow for some {positive} correlations decaying exponentially as a function of the distance between chromophores with some characteristic correlation length $r_\text{c}$ (i,e, $\kappa_{ij}=e^{-r_{ij}/r_\text{c}}$), excitons which delocalize over neighboring pigments show an increased decay rate (green and yellow solid lines).
Correlations of such form are commonly used when modelling various optical spectra of pigment-protein complexes \cite{Renger2002}.
If the correlations are negative for neighbouring pigments, the opposite behavior is observed (dashed lines).
Optical correlations between transition dipole moments give the usual fluorescence profile (red line), with few bright excitons at the bottom of the band.
We will argue later in favour of the correlations between local vibrational environments that we have postulated, by considering some experimental results.

\subsection{Other non-radiative decay pathways}

So far, we have seen how IC can be described microscopically and how it influences the non-radiative decay of different optical excitations in a molecular aggregate.
However, there could be other processes that compete with fluorescence and IC on the same timescale, which could influence the effective decay rate from a given exciton.
The other decay channel typically present in molecular systems is inter-system crossing (ISC)
\cite{Marian2011}.
During this process, a singlet excitation $S_1$ generated by optical absorption from the singlet ground state $S_0$ can be turned into an excited triplet $T_1$ by means of spin-orbit interactions.
Since the radiative transition from $T_1$ to $S_0$ is strongly suppressed, the excitation is lost non-radiatively either by IC or by triplet-triplet energy transfer to other pigments present in the molecular aggregates (i.e. carotenoids),
before it can flip the spin of a neighboring oxygen molecule, thus forming singlet oxygen, which can cause oxidative damage to the photosynthetic apparatus.

If our focus is to follow the dynamics of excitons in a narrow band (i.e. the B850 of the LH2 complex), we do not want to take into account all these processes: we are only interested in the rate at which spin-orbit interactions transfer excitations from one adiabatic PES to another.
As we explicitly worked out for IC, it has been shown that also the spin-orbit interaction results in coupling between adiabatic PESs  mediated by the nuclear momentum \cite{Englman1970}.
This tells us that the effective Hamiltonian causing ISC will exhibit couplings of the same form as \eqref{e.Hmonomer}, with properly redefined coupling constants.
The resulting ISC rate, therefore, has the same form as \eqref{e.ICrate}, i.e. it depends on the density of vibrational states at excitonic energies.
Assuming that this density of states is sufficiently flat, as done before, we have also in this case that the dependence of the rate on the excitonic state is dominated by the pattern of correlations between local baths.
ISC can be therefore absorbed together with IC into a single rate $\kappa_\alpha\Gamma_\text{non-rad}$, describing non-radiative decay of population from exciton $\alpha$.

\section{ \uppercase{Physical origin of correlated environments} }
\label{app.correl}

In this appendix we argue in favor of the presence of correlations between the vibrational environments
of single pigments belonging to the same pigment-protein complex, {which take place} on the timescale
of non-radiative decay. First, we propose an estimate based on experimental results, and later move on 
to discuss a physical mechanism which can support inter-site vibrational correlations, relying on 
inter-molecular vibrational modes.

\subsection{Phenomenological view}

Let us step back for a moment from the microscopic derivation of internal conversion, and follow another approach.
We start from experimental evidence, and use this knowledge to set up a phenomenological model of nonradiative 
decay of a molecular aggregate. The most valuable insight comes from comparing excited state lifetimes ($\tau$) 
and fluorescence quantum yields ($\phi$) of isolated pigments with those of the aggregate. Table \ref{t.qy} shows 
theses quantities in the case of bacteriochlorophyll \textit{a} molecules (\bchla) and LH2 complexes \cite{Monshouwer1997}.
Since the fluorescence quantum yield of LH2 is lower than the one of \bchla{}, there must be some additional decay 
channels in LH2 that are not present in the monomers and which can therefore be interpreted as an effect of aggregation.

\begin{table}[h]
\centering
\begin{tabular}{lcccc}
\hline
                & $\phi$ (\%) & $\tau_\text{}$ (ns)  & $\Gamma_\text{rad}$ (ns$^{-1}$) & $\Gamma_\text{non-rad}$ (ns$^{-1}$)\\
\hline
\bchla          & 18   & 3.140 & 0.057 & 0.26 \\
 LH2            & 9.86 & 0.986 & 0.10 & 0.91 \\
\hline
\end{tabular}
\caption{{\bf Single pigment vs. aggregate fluorescence.} Quantum yield of fluorescence ($\phi$) and decay lifetime ($\tau_\text{}$) of isolated \bchla{} molecules and LH2 complexes, together with their radiative ($\Gamma_\text{rad}$) and nonradiative ($\Gamma_\text{non-rad}$) decay rates. Adapted from \cite{Monshouwer1997}.}
\label{t.qy}
\end{table}

To be more quantitative, let us define the quantum yield of fluorescence
\[
\phi = \frac{\text{\# of photons emitted}}{\text{\# of photons absorbed}}
=\frac{\Gamma_\text{rad}}{\Gamma_\text{rad}+\Gamma_\text{non-rad}}
=\Gamma_\text{rad} \tau
\]
and express the radiative and nonradiative decay rates as $\Gamma_\text{rad}=\phi/\tau$ and $\Gamma_\text{non-rad}=(1-\phi)/\tau$.
The values reported in \cite{Monshouwer1997} allow us to determine radiative and nonradiative decay rates for
\bchla{} and LH2 (Table \ref{t.qy}).
While the difference in radiative lifetimes between the monomer and the aggregate is easily explained by exciton delocalization and superradiance,
a clear molecular mechanism underpinning the mismatch between nonradiative decay rates is not known with certainty.
Nonetheless, we can put forward a simple argument.
Since the nonradiative decay in LH2 is about 3.5 times faster than in \bchl{}, we can imagine the existence of additional dissipation channels.
In fact, we expect the same intra-molecular decay channels present in \bchl{} to be at play also in LH2.
Thus, the additional dissipation present in LH2 must come from some other decay pathway that has no analogue in \bchl{}, which must account for
$(\Gamma_\text{non-rad}^\text{LH2}-\Gamma_\text{non-rad}^\text{\bchl})/\Gamma_\text{non-rad}^\text{LH2}=71\%$ of the nonradiative dissipation.
The first reasonable candidate is the protein environment, which can offer other IC pathways through vibrations and conformational changes.
Since the B850 ring of LH2 is composed by dimeric subunits bound to the same protein, it makes sense to assume that the IC channels offered by the protein can in principle be correlated.
Following this argument, we allow for correlations between the vibrational environments that couple to IC and ISC transitions.
As a consequence, excitons that delocalize on neighboring pigments can experience a modified decay rate, as discussed above and shown in Fig. \ref{f.decay}.

\subsection{Microscopic mechanism}

We have seen how, within a BO framework, a coupling arises between (optical) electronic transitions and vibrations of a molecule.
A sudden change in the electronic state leads to a different electrostatic potential experienced by the nuclei, which therefore
initiate their dynamics. For this reason, this coupling involves only vibrations of the nuclei over which the electronic states
are delocalized. It is therefore clear that intra-molecular modes will experience this type of direct coupling to the electronic
dynamics.
{
However, if the chromophore is embedded in a protein environment, the vibrational modes of the protein will also influence electronic energies, leading to a direct coupling between electronic excitations of the chromophore and longer wavelength vibrational modes.
\am{This observation lies at the heart of theoretical descriptions of electronic resonances coupled to a shared phonon environment \cite{Soules1971}.}
Thus, the pure dephasing coupling in \eqref{e.Haggr} can be written explicitly in terms of these protein vibrational modes $\hat c_q$ with frequency $\nu_q$ as
\begin{equation} \label{e.HSB}
\hat{H}_\text{SB} = \sum_i \proj{i} \big[ \sum_k g_k (\hat{b}_{ik}+\hat{b}_{ik}^\dagger)
 + \sum_q \xi_{iq} (\hat{c}_{q}+\hat{c}_{q}^\dagger) \big].
\end{equation}
By redefining the operators $\hat{G}_i$ \eqref{e.G} to include also the protein modes $\hat{c}_q$, we obtain the following bath correlation functions
\begin{align}
\avg{\hat{G}_i(t) \hat{G}_j (0) }_\text{eq}
 = & \sum_{k} g_k^2 \left[ e^{-i{\omega}_k t} (n({\omega}_k) + 1 ) + e^{i{\omega}_k t}n({\omega}_k)  \right] \delta_{ij} \nonumber \\
 + & \sum_{q} \xi_{iq}\xi_{jq} \left[ e^{-i{\nu}_q t} (n({\nu}_q) + 1 ) + e^{i{\nu}_q t}n({\nu}_q)  \right].
\end{align}
While the term on the first line (corresponding to intra-molecular modes) vanishes for for $i\neq j$, the second term (arising from protein modes) is able to generate correlations between sites.
Moreover, since protein motion takes place on a larger and slower scale than intra-molecular vibrations, it is reasonable to assume that two neighboring pigments $i$ and $j$ couple to the protein motion with the same phase ($\xi_{iq}\xi_{jq}>0$). This results in positive inter-site correlations, and thus can lead to the redistribution of non-radiative decay rates discussed in \aref{app.IC}.
}
%
%
%

\begin{figure}[htb]
	\begin{center}
		\includegraphics[width=\linewidth]{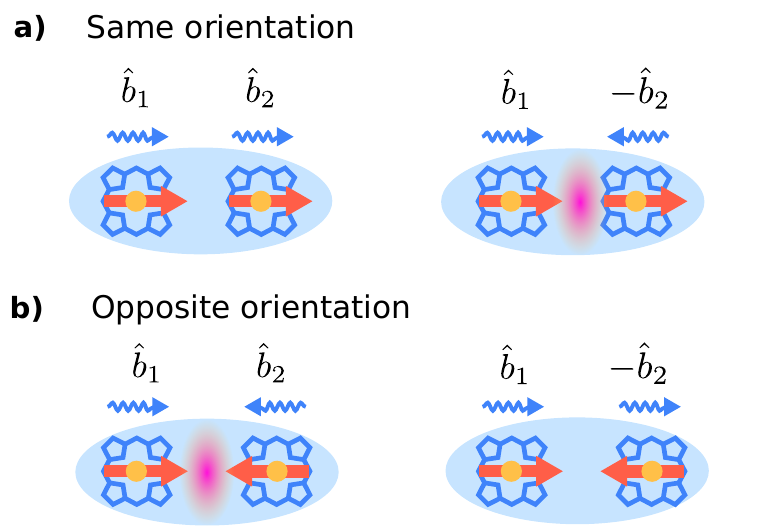}
	\end{center}
\caption{{\bf Collective modes in a molecular dimer.} (a) If the two molecules immersed in an elastic medium (roughly modeling the protein environment) have the same orientation, only the antisymmetric superposition of local vibrational modes causes streching/compression of the medium. (b) If the molecules have opposite orientation (like two neighboring B850 pigments in LH2), the symmetric superposition of local modes causes stretching/compression of the medium.}
\label{fig.spring}
\end{figure}

{Another possible mechanism able to generate positive inter-site vibrational correlations is based on the idea that the vibrational motion of a chromophore can mechanically couple to the slow vibrations of the protein environment.}
To formalize better this idea, let us consider a slightly modified version of the model described by the Hamiltonion \eqref{e.Haggr}.
To simplify the discussion, we consider a molecular aggregate composed only of two pigments, i.e. $i=1,2$,
{ and we neglect the direct coupling between electronic excitations and protein modes \eqref{e.HSB}.}
If the two pigments are
bound to the same protein, it makes sense to assume that their intra-molecular modes $\hat{b}_{1,k}$ and $\hat{b}_{2,k}$ couple to
a set of common modes $\hat{c}_q$ with frequency $\nu_q$. Since these shared modes are associated to the protein structure, it is
reasonable to assume that they are much slower and effectively classical, meaning that their energy $\nu_q$ is much smaller than
thermal energy $k_\text{B}T$. We assume that the intra-molecular modes of the two pigments couple the protein vibrations with the
same strength and phase. This choice makes sense if we think that the two pigments have opposite orientation and are embedded in
an elastic medium (see Fig. \ref{fig.spring}). The Hamiltonian of the vibrational bath then reads
\begin{align}
\hat{H}_\text{B} =& \sum_k \omega_k\left( \hat{b}_{1,k}^\dagger \hat{b}_{1,k} +
\hat{b}_{2,k}^\dagger \hat{b}_{2,k} \right) +
\sum_q \nu_q \hat{c}_q^\dagger  \hat{c}_q \\
&+ \sum_{k,q}\eta_{kq}\left(\hat{b}_{1,k}\hat{c}_q^\dagger+\hat{b}_{2,k}\hat{c}_q^\dagger + \text{ H.c.}\right), \nonumber
\end{align}
where we have neglected the counter-rotating coupling terms to simplify the following treatment, although they could be included.
Defining the symmetric and antisymmetric combinations of intra-molecular modes $\hat{b}_k = (\hat{b}_{1,k} + \hat{b}_{2,k})/\sqrt{2}$ and
${\hat{b}^{-}}_k = (\hat{b}_{1,k} - \hat{b}_{2,k})/\sqrt{2}$,
we can rewrite the coupling as
$\sum_{k,q}\sqrt{2}\eta_{kq}\hat{b}_{k}\hat{c}_q^\dagger + \text{ H.c.}$,
i.e. the antisymmetric modes ${\hat{b}^{-}}_{k}$ are insensitive to the coupling to the protein.
The symmetric modes, on the other hand, mix with the protein vibrations, yielding the new set of normal modes
\begin{align}
\hat{\tilde b}_k &= \sum_{k'} \m U_{kk'} \hat{b}_{k'} + \sum_{q} \m v_{kq} \hat{c}_q \\
\hat{\tilde c}_q &= \sum_{k} \m u_{qk} \hat{b}_k + \sum_{q'} \m V_{qq'} \hat{c}_{q'}
\end{align}
with frequencies $\tilde{\omega}_k$ and $\tilde{\nu}_q$.
Thanks to its unitarity, the transformation can be easily inverted and the bath operators $\hat G_i$ defined in \eqref{e.G} can be reexpressed as
\begin{align} \label{e.Gnormal}
\hat {G}_{1(2)} &= \frac{1}{\sqrt{2}} \left( \sum_{k} g_k\hat{b}_{k}
 \pm \sum_k g_k \hat{b}^-_k \right) + \text{ H.c.} \\
&= \frac{1}{\sqrt{2}} \left( \sum_{k} \m{G}_k \hat{\tilde b}_{k}
+ \sum_{q} \m{g}_q \hat{\tilde c}_{q} \pm \sum_k g_k \hat{b}^-_k \right) + \text{ H.c.}, \nonumber
\end{align}
where we have defined the transformed system-bath couplings
\begin{equation}
    \m{G}_k = \sum_{k'} g_{k'} \m{U}^*_{k'k}
    \text{\qquad and \qquad}
    \m{g}_q = \sum_{k'} g_{k'} \m{u}^*_{k'q}.
\end{equation}
Analogous expressions can be found for the operators $\hat {F}_i$ defined in \eqref{e.F}.

Let us now focus on the bath correlation functions for the pure dephasing environment introduced in Eq. \eqref{e.GGeq}, evaluated on the thermal state $\rho_\text{eq}\propto e^{-\hat H_\text{B}/k_\text{B}T}$. (The system-bath coupling leading to internal conversion can be treated in a completely analogous way.)
Using the new normal mode decomposition of $\hat{G}_i$ \eqref{e.Gnormal}, we get
\begin{multline} \label{e.G1G2}
\avg{\hat{G}_1(t)\hat{G}_{1(2)}(0)}_\text{eq}=
\frac{1}{2} \sum_{k} \m{G}_k^2 \left[ e^{-i\tilde{\omega}_k t} (n(\tilde{\omega}_k) + 1 ) + e^{i\tilde{\omega}_k t} n(\tilde{\omega}_k)  \right] \\
\quad
+\frac{1}{2}\sum_{q} \m{g}_q^2 \left[ e^{-i\tilde{\nu}_k t} (n(\tilde{\nu}_q) + 1 ) + e^{i\tilde{\nu}_q t} n(\tilde{\nu}_q)  \right] \\
\pm \frac{1}{2}\sum_{k} g_k^2 \left[ e^{-i{\omega}_k t} (n({\omega}_k) + 1 ) + e^{i{\omega}_k t}n({\omega}_k)  \right].
\end{multline}
To lowest order in the inter-site vibrational coupling $\eta_{kq}$, we find the following perturbative expressions
\begin{align}
\m{G}_k &\approx g_k, &
\m{g}_q&\approx-\sum_k \frac{\sqrt{2} g_k \eta_{kq}}{\omega_k-\nu_q}, \\
\tilde \omega_k &\approx\omega_k + \sum_q \frac{2\eta_{kq}^2}{\omega_k-\nu_q}, &
\tilde \nu_q &\approx\nu_q - \sum_k \frac{2\eta_{kq}^2}{\omega_k-\nu_q}.
\end{align}
Note that, in order to be perfectly consistent, we should keep also the second-order correction to $\m{G}_k$.
However, since the protein modes generally have much smaller energies $\nu_q$ than both $k_\text{B}T$ and  intra-molecular modes $\omega_k$, their thermal occupation number is much higher, i.e. $n(\tilde \nu_q)\gg n(\tilde  \omega_k)$.
Therefore, the correction arising from the first line in \eqref{e.G1G2} is negligible with respect to the one originating from the second line.
Taking into account all these assumptions, we obtain
\begin{align}
\avg{\hat{G}_1(t)\hat{G}_{1}(0)}_\text{eq} & \approx \sum_{k} g_k^2 \left[ e^{-i{\omega}_k t} (n({\omega}_k) + 1 ) + e^{i{\omega}_k t}n({\omega}_k)  \right] \label{e.11} \\
\avg{\hat{G}_1(t)\hat{G}_{2}(0)}_\text{eq} & \approx \sum_q \left(\sum_{k} \frac{g_k \eta_{kq}}{\omega_k - \nu_q} \right)^2 \label{e.12} \\
 & \qquad \times \left[ e^{-i{\nu}_q t} (n({\nu}_q) + 1 ) + e^{i{\nu}_q t}n({\nu}_q) \right]. \nonumber
\end{align}
Note that both these equations can be rewritten in terms of positive spectral densities:
In the case of \eqref{e.11} it is the one defined in \eqref{e.SD},
whereas for \eqref{e.12} we can define
\begin{equation}
\Delta\mathcal{J}(\omega) = \sum_q \left(\sum_{k} \frac{g_k \eta_{kq}}{\omega_k - \nu_q} \right)^2 \delta(\omega-\nu_q).
\end{equation}
Taking the Fourier transform of these correlation functions, we obtain the rates that allow us to write down a Lindblad equation for the reduced dynamics of the excitons, which are proportional to the spectral densities $\mathcal{J}(\omega)$ and $\Delta\mathcal{J}(\omega)$.

At this point, we can see that our particular choice of coupling the local intra-molecular modes to the common protein modes with the same phase leads to the establishment of positive inter-site correlations ($\Delta\mathcal{J}(\omega)>0$).
If we chose to couple local modes with opposite phase instead, we would end up with negative correlations between the vibrational environments.

\section{ \uppercase{ Lineshapes, linear spectra and and transfer rates } } \label{app.lineshapes}

In this section we determine the lineshapes of an excitonic system (i.e. our unit cells), which allow for the calculation of both linear optical spectra and
excitonic transfer rates.
Consider an excitonic system with Hamiltonian $\hat{H}=\sum_{\alpha}E_\alpha\proj{\alpha}$, which we identify as our unit cell,
where Greek letters denote exciton states of the unit cell $\ket{\alpha}=\sum_i\ket{i} \braket{i}{\alpha}$. (Here we relax the convention used in the main text, according to which primed indices refer to the same subunit.)
Within a single unit cell, excited state population thermalizes due to the interaction between electronic and vibrational degrees of freedom of both intra- and inter-molecular origin.
The main electron-phonon coupling mechanism is described by the pure-dephasing interaction
presented in the aggregate Hamiltonian \eqref{e.Haggr}.
Under the conditions of weak electron-phonon coupling and fast vibrational relaxation, this process can be described by a Lindblad equation.
In the presence of correlations $\kappa_{ij}$ between different local environments,
we can write down the resulting Lindblad dissipator as
\begin{align} \label{e.Drel}
\mathcal{D} \rho =& \sum_{\alpha\neq\beta} R_{\alpha\beta} \left( \ket{\alpha}\bra{\beta} \rho \ket{\beta}\bra{\alpha} - \frac{1}{2} \{ \proj{\beta}, \rho \} \right) \\
 & + 2\gamma \sum_{\alpha,\beta} \zeta_{\alpha\beta} \left( \proj{\alpha}\rho\proj{\beta}
 -\frac{1}{2}\delta_{\alpha\beta}\{\proj{\alpha}, \rho \} \right),
\nonumber
\end{align}
where the first line describes population transfer across different excitons and the second describes pure dephasing processes.
The rates can be obtained respectively as
\begin{align}
R_{\alpha\beta} &= 2\pi \mathcal{J}(|\omega_{\alpha\beta}|)|n(\omega_{\alpha\beta})| \sum_{i,j} \braket{\alpha}{i}\braket{i}{\beta} \kappa_{ij} \braket{\beta}{j}\braket{j}{\alpha}, \label{e.Rab} \\
2\gamma\zeta_{\alpha\beta} &= 2\gamma \sum_{i,j} \braket{\alpha}{i}\braket{i}{\alpha} \kappa_{ij} \braket{\beta}{j}\braket{j}{\beta}, \label{e.gab}
\end{align}
where
$\omega_{\alpha\beta}=E_\alpha-E_\beta$, and $\gamma$ is the single pigment optical dephasing rate.
Note that, when the local vibrational environments are not correlated, i.e. $\kappa_{ij}=\delta_{ij}$, the factors involving excitonic amplitudes in \eqref{e.Rab} and \eqref{e.gab} reduce to the spatial overlap between excitons and their participation ratio.
Lindblad dynamics in the exciton basis results  in an exponential decay of the optical coherences between exciton $\alpha$ and the ground state $g$ with a rate
$\gamma_\alpha = \gamma\zeta_{\alpha\alpha} + \sum_{\beta (\neq\alpha)}R_{\beta\alpha}/2$, whereas the inter-exciton coherence decays
with a rate
$\gamma_{\alpha\beta} = \gamma_\alpha + \gamma_\beta - 2\gamma \zeta_{\alpha\beta}$.
This leads to simple expressions when computing optical absorption and emission spectra.
These are related to the absorption and emission tensors,
$\m{A}(\omega)$ and $\m{E}(\omega)$ respectively,
defined by their matrix elements
\begin{align} \label{e.AE}
\m{A}_{\alpha\beta}(\omega) &= \int \text{d}t e^{i\omega t} \langle \hat{\sigma}_{\alpha}(t) \hat{\sigma}_{\beta}^{\dagger}(0) \rangle_g \nonumber \\
\m{E}_{\alpha\beta}(\omega) &= \int \text{d}t e^{-i\omega t} \langle \hat{\sigma}_{\beta}^{\dagger}(t) \hat{\sigma}_{\alpha}(0) \rangle_e.
\end{align}
Here $\hat{\sigma}_{\mu}(t)$ denotes the Heisenberg time evolution of the annihilation operator of exciton
$\mu$, $\hat{\sigma}_{\mu} = \ket{g}\bra{\mu}$,
and $\langle\cdot\rangle_{g (e)}$ denotes the average over the equilibrium ground (excited) state.
Evolving the transition dipole operator through the dual of $\mathcal{D}$,
results in the simple expression $\hat{\sigma}_{\mu}(t) = \ket{g}\bra{\mu} e^{-(\gamma_{\mu}+iE_{\mu})t}$,
which leads to diagonal absorption and emission tensors, with each exciton having a Lorentzian lineshape, i.e.
$\m{A}_{\alpha\beta}(\omega) = \delta_{\alpha\beta} f_{\alpha}(\omega)$ and
$\m{E}_{\alpha\beta}(\omega) = \delta_{\alpha\beta} f_{\alpha}(\omega) p\super{th}_{\alpha}$,
where $p\super{th}_{\alpha}$ is the thermal population of exciton $\alpha$, and
\begin{equation} \label{e.AEdeph}
f_{\alpha}(\omega) = \frac{2 \gamma_{\alpha}}{\gamma_{\alpha}^2 + (\omega-E_{\alpha})^2}.
\end{equation}
Once we have the lineshape for each exciton, we weight each individual lineshape by its associated brightness $\chi_{\alpha} \approx |\mathbf{D}_{\alpha}|^2/d^2$, and calculate straightforwardly the absorption spectrum of the unit cell as
$\omega \sum_\alpha \chi_\alpha f_{\alpha}(\omega)$.
In the case of the B850 subunit, once averaged over static disorder, this expression gives excellent agreement with experimental results, as shown in the main text (\fref{fig2}{B}).

If we introduce a second excitonic system, weakly interacting (with respect to the timescales associated to dephasing) with the first one via dipole-dipole couplings
\begin{equation}
V_{ii'}=\matrel{i}{\hat V}{i'} = \frac{1}{4\pi\epsilon_0}\frac{d^2}{r_{ii'}^3}
\left[ \mathbf{e}_i \cdot \mathbf{e}_{i'} - 3 (\mathbf{e}_i \cdot \mathbf{n}_{ii'})  (\mathbf{e}_{i'} \cdot \mathbf{n}_{ii'}) \right],
\end{equation}
we can calculate the incoherent energy transfer rate between excitons $\alpha$ and $\alpha'$ on the two subunits via generalized F\"orster theory, according to which, we have
\begin{align} \label{e.W}
W_{\alpha\alpha'}&=\frac{|V_{\alpha\alpha'}|^2}{2\pi} \int\mathrm{d}\omega f_{\alpha}(\omega)
f_{\alpha'}(\omega) \nonumber \\
 &= {2|V_{\alpha\alpha'}|^2}\ \frac{\gamma_{\alpha\alpha'}}{\omega_{\alpha\alpha'}^2+\gamma_{\alpha\alpha'}^2},
\end{align}
where $V_{\alpha\alpha'} = \sum_{i,i'} \braket{\alpha}{i}V_{ii'}\braket{i'}{\alpha'}$,
\am{which reduces to \eref{e.kpm} from the main text in the case of dimeric unit cells, once $\hbar$ is reintroduced.}

\section{ \uppercase{ Parametrization of the B850 ring } }
\label{app.B850}

\begin{figure}[htb]
	\begin{center}
		\includegraphics[width=\linewidth]{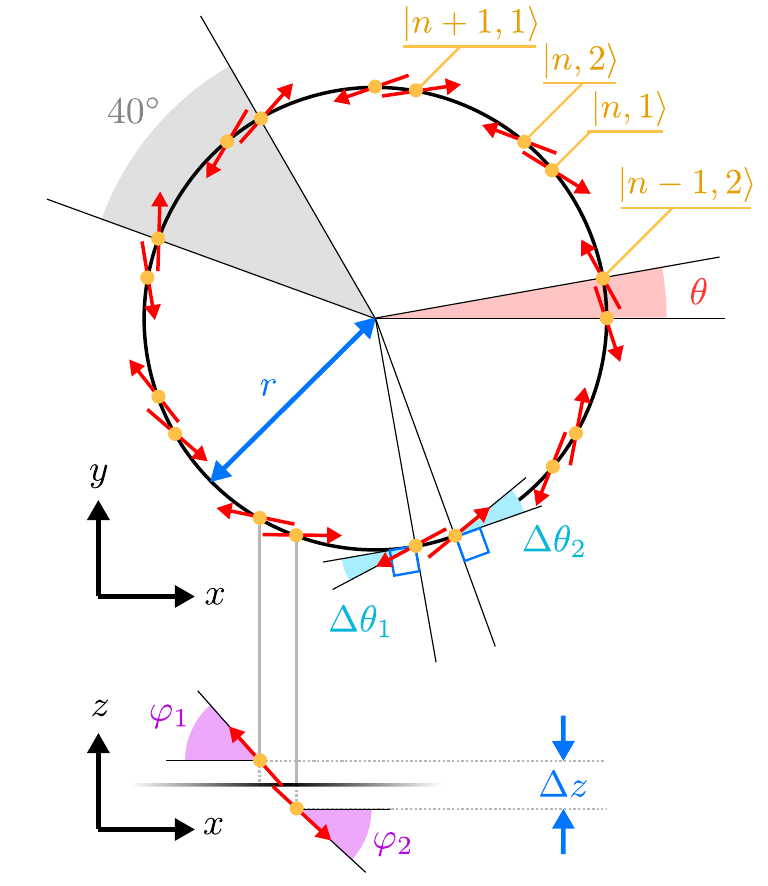}
	\end{center}
\caption{{\bf Geometry of a B850 ring.} Projections on the $xy$-plane (top) and on the $xz$-plane (bottom) of a B850 ring lying on the $xy$-plane. The $z$-direction is magnified by a factor 10 and the angles $\varphi_1$ and $\varphi_2$ are rescaled accordingly.}
\label{fig.lh2}
\end{figure}

\fref{fig.lh2}{} and Table \ref{t.lh2} summarize the geometry and the parameters that have been used in the simulations of the B850 rings.
The ring structure is dimerized, meaning that each pigment $i$ is identified by two indices, one specifying the dimer $(n=1,\dots,9)$ and the other the position within the dimer $(\nu=1,2)$.
The two bacteriochlorophyll (\bchl) molecules belonging to the same dimer are usually labelled $\alpha$ and $\beta$.
A single ring is described by the Hamiltonian
\begin{align}
\hat{H} & = \sum_{i} (\varepsilon +\delta\varepsilon_{i}) \ket{i}\bra{i}+\sum_{i\ne j}^{} J_{ij}\ket{i}\bra{j} \nonumber \\
& =
\sum_{n=1}^{9} \big[
  (\varepsilon +\delta\varepsilon_{n,1}) \proj{n,1}
+ (\varepsilon +\delta\varepsilon_{n,2}) \proj{n,2} \nonumber  \\
& + J_1( \ket{n,1 }\bra{n,2} + \text{ H.c.}   )
  + J_2( \ket{n,2 }\bra{n+1,1} + \text{ H.c.} ) \big] \nonumber \\
& + \sum_{i,j}^{}{\vphantom{\sum}}'  J_{ij} \ket{i}\bra{j} 
\end{align}
where the primed sum indicates summation over all couples of non-adjacent pigments and $J_{ij}$ stands in this case for the  dipole-dipole interaction between pigments.
The fluctuations of the site energies $\delta\varepsilon_{i}$ are given by the sum of two Gaussian random variables: one with standard deviation $\sigma_\text{p}$, completely uncorrelated for different pigments, describing local energy shifts due to the slightly different protein environments; one with standard deviation $\sigma_0$, which is the same for all pigments, describing global shifts of the ground state energy.

\begin{table}[htb]
	\begin{center}
		\caption{Model parameters for a B850 ring.}
		\begin{tabular}{C{0.12\linewidth}C{0.3\linewidth}L{0.55\linewidth}}
			\hline \hline
			& Value & Parameter description \\
			\hline \hline
			$\theta$      & $10.06^{\circ}$               & intra-dimer angle \am{\cite{Caycedo2017}} \\	\hline
			$\Delta\theta_{1}$, $\Delta\theta_{2}$  & $19.9^{\circ}$, $17.6^{\circ}$ & dipole tilt away from the tangent \am{\cite{Caycedo2017}} \\	\hline
			$\varphi_{1}$, $\varphi_{2}$                  & $6.24^{\circ}$, $4.81^{\circ}$ & dipole tilt away from ring plane \am{\cite{Caycedo2017}} \\	\hline
			$r$           & \textrm{3.0 nm}               & ring radius \am{\cite{Walz1998}} \\	\hline
			$\Delta z$    & \textrm{0.063 nm}             & vertical displacement between $\alpha$ and $\beta$ \bchl{} \am{\cite{Caycedo2017}} \\	\hline
			$d$           & \textrm{6.4 D}                & $Q_y$ dipole moment of \bchl{} \am{\cite{Pullerits1997}} \\	\hline
			$\varepsilon$ & \textrm{12330 cm$^{-1}$}      & average site energy \am{\cite{Caycedo2018}} \\ \hline
			$J_1$, $J_2$  & \textrm{320, 255 cm$^{-1}$}   & intra- and inter-dimer nearest neighbor coupling \am{\cite{Scholes1999}} \\ \hline
			$\gamma$      & \textrm{(20 fs)$^{-1}$}       & single pigment optical dephasing rate \am{\cite{Caycedo2018}} \\ \hline
			$R_{01}$      & \textrm{(200 fs)$^{-1}$}      & relaxation rate between the two lowest excitons \am{\cite{Stuart2011}}\\ \hline
			$\Gamma_\text{th}$   & \textrm{(1 ns)$^{-1}$} & decay rate from thermalized exciton manifold \am{\cite{Monshouwer1997}} \\ \hline			
			$T$           & \textrm{298 K}                & temperature of the vibrational bath \\ \hline
			$\phi$        & 0.1 & quantum yield of fluorescence \am{\cite{Monshouwer1997}} \\	\hline
			$\sigma_\text{p}$, $\sigma_\text{0}$ & 265 cm$^{-1}$, 33 cm$^{-1}$ & std. dev. of uncorrelated/ground state disorder \am{\cite{Caycedo2018}}\footnote{\am{These values have been rescaled from \cite{Caycedo2018} to match the B850 absorption linewidth.}} \\	\hline
			$r_\text{c}$  & 0.5 nm                        & correlation radius of local pigment environments \am{\cite{Renger2002}} \\				
			\hline \hline		
		\end{tabular}
		\label{t.lh2}
	\end{center}
\end{table}

{Throughout this work, the B800 ring enters the dynamics only during the excitation process as we describe more in detail 
in the next section.} However, we explicitly included B800 rings  in the calculation of the absorption spectra shown in \fref{fig2}{b}, 
for a better comparison with the experimental data. We model the B800 ring as a set of $9$ transition dipoles arranged on a circle 
of radius $3.1$ nm, concentric to the B850 ring and vertically displaced from it by $1.7$ nm. We consider the B800 dipoles to be 
perfectly coplanar and tangent to the circle. In accordance with previous works \cite{Caycedo2018}, we set the ratio between 
B850 to B800 pigment dipole strength to $1.1$. The site energy of B800 pigments $\varepsilon_\text{B800}$ has an average value 
of $1.25\ 10^{4}$~cm$^{-1}$, with Gaussian static disorder with standard deviation $\sigma_\text{B800}=100$~cm$^{-1}$, whereas 
the single pigment optical dephasing rate is $\gamma_\text{B800}=70$~cm$^{-1}$ \am{\cite{Caycedo2018}}.

\begin{figure}[htb]
	\begin{center}
		\includegraphics[width=0.7\linewidth]{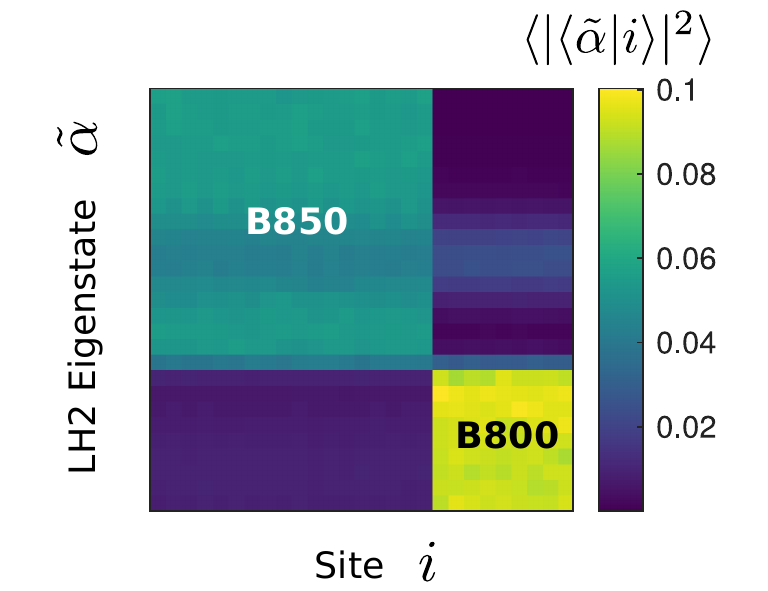}
	\end{center}
\caption{
{\bf B800-B850 mixing in a LH2 ring.}
Disorder-averaged populations of LH2 mixed B800-B850 eigenstates $\ket{\tilde \alpha}$ on different sites $\ket{i}$.
The eigenstates are clearly divided in two blocks, representing the fact that there is negligible coherent mixing between B800 and B850 excitons. The ensemble average is obtained from 10$^4$ realizations of static disorder.}
\label{fig.lh2_states}
\end{figure}

The B800 pigments couple to each other and to the B850 pigments via dipole-dipole interactions. As a result, {one may expect
that} excitons can in principle delocalize across both rings. However, the presence of static disorder is sufficient to destroy 
any inter-ring delocalization {\cite{Caycedo2018}}. This can be seen in \fref{fig.lh2_states}, where we plot the average population of a LH2 exciton 
$\ket{\tilde{\alpha}}$ (diagonalizing the total B800 + B850 Hamiltonian) on site $\ket{i}$. The absence of significant B800-B850 
coherent mixing thus justifies the approach adopted throughout the paper, where we only consider the indirect effect of incoherent 
B800-to-B850 energy transfer {to populate the B850 manifold after initial laser excitation}.


\section{ \uppercase{Simulations of a LH2 array} }
\label{app.chain}

\begin{figure}[htb]
	\begin{center}
		\includegraphics[width=\linewidth]{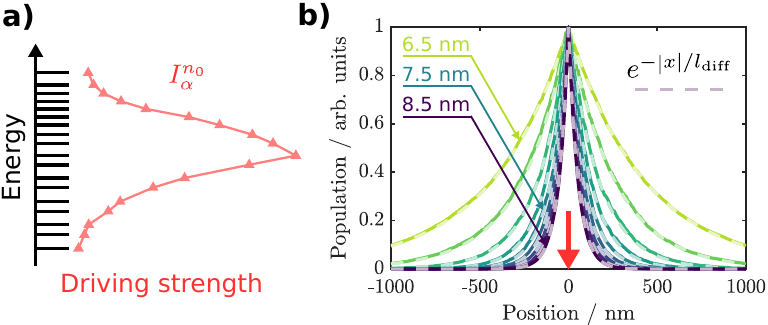}
	\end{center}
	\caption{
		\ftit{Driving profile and excitonic stationary spatial distribution.}
		\subf{a}
		Driving at the central ring takes place from the B800 pigments,
		with a distribution across the excitonic spectrum given by \eqref{inj}. The profile is averaged over $10^4$ realizations of static disorder.
		\subf{b} Injection at a single site (at $x=0$) results in an exponential exciton distribution along the light-harvesting array.
		Results are shown for inter-complex distances ranging from 6.5 nm to 8.5 nm.
		The exciton distributions obtained numerically (solid lines) fully agree with exponential distributions of different diffusion length (dashed lines).
		Numerical results are obtained as averages over $10^3$ realizations of static disorder.
	}
	\label{fig8}
\end{figure}

\begin{figure}[htb]
	\begin{center}
		\includegraphics[width=\linewidth]{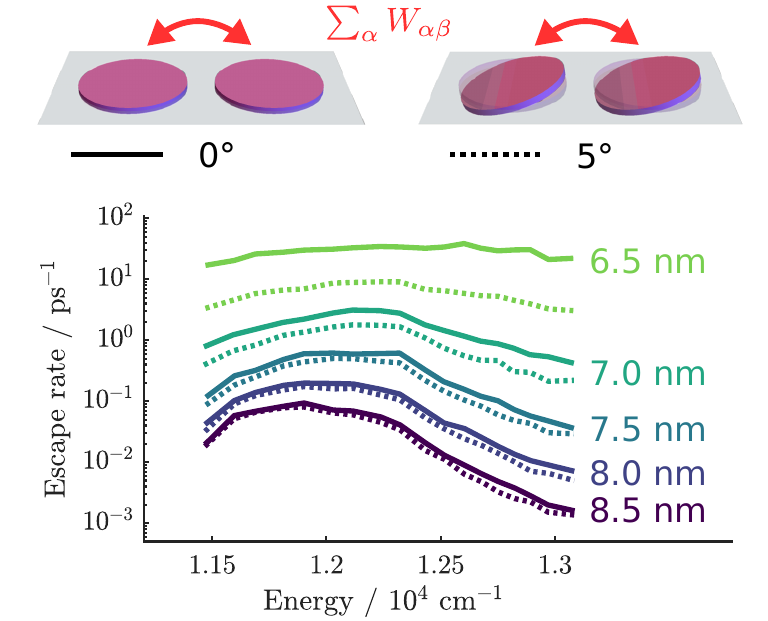}
	\end{center}
	\caption{
		\ftit{Energy transfer between tilted B850 rings.} Total energy transfer rate from exciton $\beta$ to any other exciton $\alpha$ on a neighboring ring (escape rate), as a function of the energy of the initial excitonic state $\beta$. Results for coplanar (tilded) rings are shown with solid (dotted) lines, for inter-ring distances ranging from 6.5 nm (green) to 8.5 nm (purple). Even a slight tilt of $5^\circ$ outside the plane results to a significant reduction of the transfer rate at short inter-ring distance. Results are obtained as averages over $10^3$ realizations of static disorder.
	}
	\label{fig_tilt}
\end{figure}

In this section we give more details on the simulations of the full LH2 linear array.
In analogy with the experiment of Escalante \textit{et al.} \cite{Escalante2010},
we look at the stationary exciton probability profile along the linear array upon continuous wave driving, i.e. we look numerically for the  steady state of \eref{Pauli} (with all rates dependent also on the specific subunit $n$, since every subunit exhibits a different realization of static disorder, i.e. different spectra and pigment positions, affecting relaxation, fluorescence, transfer and injection rates).
In the experiment the driving is provided by a 800 nm laser with spatial intensity profile with a full width at half maximum of $400\pm 50$ nm.
In order to simplify the numerics, we consider instead injection on a single B850 complex at the center of the chain.
In this way, we can simulate shorter chains (up to 1001 subunits) and be safely protected from systematic errors introduced by the finite size of the array.
The results for spatially broad excitation can be recovered as shown in \aref{app.toymodel} in the case  of the {exactly solvable} model, i.e. by convolving the result for local injection with the desired excitation profile.

We clarified that our injection profile is local, i.e. $I^{n}_{\alpha}\propto\delta_{n_0 n}$ where $n_0=501$ is the central site of the chain, but we have not specified yet in which states $\alpha$ of the ring excitations are injected.
Optical excitation at 800 nm cannot be absorbed by the B850 subunit (as seen in \fref{fig2}{B}), nevertheless it  can excite the B800 ring.
We are not explicitly considering the B800  subunit in our model.
However, excitations  enter the B850 ring upon downhill B800 $\to$ B850 energy transfer.
Due to the small coherent coupling between the two concentric rings,
this transfer process is largely controlled by incoherent rates
of the form of \eref{e.W},
therefore we can think of an indirect excitation of the B850 excitons with an energy distribution given by
\begin{equation} \label{inj}
	I^{n_0}_\alpha \propto \frac{(\gamma_{\alpha}+\gamma\sub{B800})}{(E_\alpha-\varepsilon\sub{B800})^2+(\gamma_{\alpha}+\gamma\sub{B800})^2},
\end{equation}
where
{
$\varepsilon\sub{B800}$ and $\gamma\sub{B800}$
are optical gap and dephasing rate of the B800 pigments,
whose values are given in \aref{app.B850}.
}
The disorder-averaged injection profile \eqref{inj} is shown in \fref{fig8}{a}.

The resulting stationary probability profiles shown in \fref{fig8}{b} for different lattice steps $l$ ranging from 8.5 nm to 6.5 nm, after averaging over $10^3$ realizations of static disorder.
The probabilities $\bar{p}^n_{\alpha}$, shown as a function of the position $x=(n-n_0)l$,
exhibit  a clear exponential decay around the injection site in the middle of the array.
A convolution of these exciton population profiles with a Gaussian injection profile with 400 nm  full width at half maximum yields the distributions reported in \fref{fig3}{b}.

All simulations are performed for coplanar arrangements of B850 rings.
As discussed in the main text, the nano-engineered arrays analyzed here allow for coplanar B850 rings.
However, in the main text we also note that in biological light-harvesting membranes, an angle of about 5$^\circ$ relative to the aggregation plane is observed.
This small tilt leads to slower energy transfer, especially for short inter-ring distances, as shown in \fref{fig_tilt}{}.

\am{
Lastly, we note that all simulated LH2 arrays feature a fixed inter-complex distance $l$ throughout the length of the array. This clearly represents an approximation, since every realistic macromolecular assembly will show fluctuations in the lattice constant. While these large-scale geometric defects are expected to severely limit the range of ballistic energy transfer and eventually lead to diffusive transport, they only act as an additional source of static disorder in our model, which already describes energy transfer in the diffusive regime.
Thus, in our case, this further noise source will lead to changes in the transfer rates but will not affect the scaling of diffusion length with respect to the lattice constant, leaving our main conclusions unaltered.
Nevertheless, the effect of lattice constant fluctuations can be estimated by analyzing the dependence of the diffusion length $l_\text{diff}$ on the inter-complex distance $l$, shown in \fref{fig3}{d}.
This seems to be  described by a monotonically decreasing convex function. In other words, this means that the increase in diffusion length that we obtain by shortening the lattice constant $l$ by an amount $\delta l$ exceeds the decrease that we would obtain by expanding the lattice constant by the same amount to $l+\delta l$. Thus, when assuming a symmetric distribution of lattice constants around $l$, the averaging would lead to a slightly larger diffusion length.
However, we note that this effect might be too small to be observed in our case, due to the presence of other sources of static disorder.
}


\bibliographystyle{apsrev4-1}

\end{document}